\global\def\draftcontrol{0}

 \def\versionno{ cdsw -- version 1.0 }
\catcode`\@=11

\expandafter\ifx\csname draftcontrol\endcsname\relax\global\def\draftcontrol{0}
\fi
{\count255=\time\divide\count255 by 60
\xdef\hourmin{\number\count255}
\multiply\count255 by-60\advance\count255 by\time
\xdef\hourmin{\hourmin:\ifnum\count255<10 0\fi\the\count255}}
\def\draftdate{\number\month/\number\day/\number\year\ \ \ \hourmin }

\newcommand\makepapertitle{\par
  \begingroup
    \renewcommand\thefootnote{\@fnsymbol\c@footnote}%
    \def\@makefnmark{\rlap{\@textsuperscript{\normalfont\@thefnmark}}}%
    \long\def\@makefntext##1{\parindent 1em\noindent
            \hb@xt@1.8em{%
                \hss\@textsuperscript{\normalfont\@thefnmark}}##1}%
     \newpage
     \global\@topnum\z@   % Prevents figures from going at top of page.
     \@makepapertitle
     \thispagestyle{empty}\@thanks
  \endgroup
  \setcounter{footnote}{0}%
  \global\let\thanks\relax
  \global\let\makepapertitle\relax
  \global\let\@makepapertitle\relax
  \global\let\@thanks\@empty
  \global\let\@author\@empty
  \global\let\@date\@empty
  \global\let\@title\@empty
  \global\let\title\relax
  \global\let\author\relax
  \global\let\date\relax
  \global\let\and\relax
  \def\version{\let\version\@version\@gobble}
}
\def\@makepapertitle{%
  \newpage
   \ifnum\draftcontrol=1 {}
   \version\versionno
   \vskip 3em%
   \else
   \hfill\hbox to 3cm {\parbox{4cm}{\@pubnum}\hss}%
   \vskip 3em%
   \fi
   \begin{center}%
   \let \footnote \thanks
     {\LARGE \@title \par}%
     \vskip 1.5em%
     {\normalsize%\large
       \lineskip .5em%
       \begin{center} %\begin{tabular}[t]{c}%
         \@author
       \end{center} % \end{tabular}
\par}%
     \vskip 1em%
     {\@bstract}%
     \end{center}%
     \vskip .5em
     \@date%
   \par
}

\gdef\@pubnum{}
\def\pubnum#1{%
  \gdef\@pubnum{#1}}

\gdef\@bstract{}
\def\Abstract#1{%
  \gdef\@bstract{%
   \parbox{\textwidth-0pc}{%
   \centerline{\bf Abstract}%\penalty1000
   \noindent%\abstractfont \baselineskip=12pt
   \renewcommand\baselinestretch{1.0}
   {#1}}}
}

\def\ps@paper{\let\@mkboth\@gobbletwo%
     \ifnum\draftcontrol=1
        \def\@oddfoot{\hbox to \textwidth{\tiny \versionno \hfil\tiny\draftdate}%
        \hskip -\textwidth \hbox to \textwidth{\hfil\rm\thepage\hfil}}%
     \else\def\@oddfoot{\hbox to \textwidth{\hfil\rm\thepage\hfil}}
     \fi
     \let\@evenfoot\@oddfoot
}
\newenvironment{acknowledgments}{%
\vskip 3.25ex
\noindent {\bf Acknowledgments}
}

\def\@version#1{\ifnum\draftcontrol=1
\typeout{}\typeout{#1}\typeout{}
\vskip3mm\centerline{\hbox{\fbox{\normalsize{\tt DRAFT -- #1 -- }
                   {\draftdate}}}}\vskip3mm
\fi}
\let\version\@version
\long\def\eqlabel#1{\ifnum\draftcontrol=1
                    \tag@false  % there are some problems with multline without this
                    \tag*{(\theequation) \hbox to -0.2cm{\hspace{0cm}\small{#1}\hss}}
                    \refstepcounter{equation} 
                    \edef\@currentlabel{\theequation}
                    \ltx@label{#1}          % use old LaTeX \label instead of new definition
                    \else
                    \label{#1}
                    \fi
                    }
\let\st@bibitem\@bibitem
\let\st@lbibitem\@lbibitem
\ifnum\draftcontrol=1
  \def\@bibitem#1{%
    \st@bibitem{#1}\a@@label{#1}\ignorespaces}
  \def\@lbibitem[#1]#2{%
    \st@lbibitem[#1]{#2}\a@@label{#2}\ignorespaces}
  \def\a@@label#1{%
    \gdef\a@lab{\smash{\normalfont\small#1}}
    \ifvmode
      \if@inlabel
        \global\setbox\@labels\hbox{%
          \llap{\a@lab\let\a@lab\relax
                \kern\@totalleftmargin\kern\marginparsep}%
          \box\@labels}%
      \fi
    \fi}
\fi
\documentclass[11pt,twoside]{article}
\usepackage{amsmath,amssymb,array,calc,epsfig,psfrag,atmp04}
\begin{document}
%%%
%%%%%% relax
%%%%%%%%%
\ifnum\draftcontrol=1
\tolerance=1000
\fi
%%%%%%%%%%%%

%%%
%%%%%% layout
%%%%%%%%%
%\addtolength{\oddsidemargin}{0.5cm}
%\addtolength{\evensidemargin}{-0.5cm}
%\renewcommand\baselinestretch{1.25}
%\setlength{\paperheight}{11in}
%\setlength{\paperwidth}{8.5in}
%\setlength{\textwidth}{\paperwidth-3.5in}     \hoffset= -.3in   % +1in from printer
%\setlength{\textheight}{\paperheight-2.6in}   \topmargin= -.6in % +1in from printer

%%%%%%%%% section titles
\renewcommand\section{\@startsection {section}{1}{\z@}%
                                   {-3.5ex \@plus -1ex \@minus -.2ex}%
                                   {2.3ex \@plus.2ex}%
                                   {\normalfont\large\bfseries}}
\renewcommand\subsection{\@startsection{subsection}{2}{\z@}%
                                     {-3.25ex\@plus -1ex \@minus -.2ex}%
                                     {1.5ex \@plus .2ex}%
                                     {\normalfont\normalsize\bfseries}}
\renewcommand\subsubsection{\@startsection{subsubsection}{3}{\z@}%
                                     {-3.25ex\@plus -1ex \@minus -.2ex}%
                                     {1.5ex \@plus .2ex}%
                                     {\normalfont\normalsize\it}}

%%%
%%%%%% number equations within sections
%%%%%%%%% 
\numberwithin{equation}{section}

%%%
%%%%%% macros 
%%%%%%%%%

%%%%%%%%% standard
%%%%%%%%%%%%

\def\cala         {{\cal A}}
\def\calA         {{\mathfrak A}}
\def\calAbar      {{\underline \calA}}
\def\calb         {{\cal B}}
\def\calc         {{\cal C}}
\def\cald         {{\cal D}}
\def\cale         {{\cal E}}
\def\calf         {{\cal F}}
\def\calg         {{\cal G}}
\def\calG         {{\mathfrak G}}
\def\calh         {{\cal H}}
\def\cali         {{\cal I}}
\def\calj         {{\cal J}}
\def\calk         {{\cal K}}
\def\call         {{\cal L}}
\def\calm         {{\cal M}}
\def\caln         {{\cal N}}
\def\calo         {{\cal O}}
\def\calp         {{\cal P}}
\def\calq         {{\cal Q}}
\def\calr         {{\cal R}}
\def\cals         {{\cal S}}
\def\calt         {{\cal T}}
\def\calu         {{\cal U}}
\def\calv         {{\cal V}}
\def\calw         {{\cal W}}

\def\complex      {{\mathbb C}}
\def\naturals     {{\mathbb N}}
\def\projective   {{\mathbb P}}
\def\rationals    {{\mathbb Q}}
\def\reals        {{\mathbb R}}
\def\zet          {{\mathbb Z}}

\def\del          {\partial}
\def\delbar       {\bar\partial}
\def\ee           {{\rm e}}
\def\ii           {{\rm i}}
\def\chain        {{\circ}}

\def\ie{{\it i.e.}}
\def\eg{{\it e.g.}}

\def\revise#1       {\marginpar{\rule{2mm}{1cm} #1}}
\newcommand\fnxt[1] {\raisebox{.12em}{\rule{.35em}{.35em}}\mbox{\hspace{0.6em}}#1}
\newcommand\nxt[1]  {\\\fnxt#1}

%%%%%%%%% paper specific macros
%%%%%%%%%%%%

\def\ZZ{\zet}
\def\RR{\reals}
\def\PP{\projective}
\def\RP{\RR\PP}
\def\N{\caln}

\def\phir{\undertilde{\phi}}
\def\phii{\Im(\phi)}
\def\zbar{{\bar z}}
\def\zr{{\zeta}}
\def\ybar{{\bar y}}
\def\yr{{\upsilon}}
\def\xbar{{\bar x}}
\def\xr{{\xi}}
\def\Wr{{\calw}}
\def\Wbar{{\overline W}}
\def\Ar{{\undertilde A}}
\def\Dr{{\undertilde D}}
\def\Er{{\undertilde E}}
\def\L{{\rm L}}
\def\R{{\rm R}}
\def\Phibar{{\bar \Phi}}
\def\chibar{{\overline\chi}}
\def\omegat{{\tilde\omega}}
\def\de#1#2{{\rm d}^{#1}\!#2\,}
\def\De#1{{\cald}#1\,}
\def\undertilde#1{{\vphantom#1\smash{\underset{\widetilde{\hphantom{\displaystyle#1}}}{#1}}}}
\def\twin{{{\rm tr}_{\rm tw}(-1)^F}}
\def\unin{{{\rm tr}_{\rm un}(-1)^F}}
\def\Wilson{\Sigma}
\def\onecycle{\gamma}

\def\fus{\star}
\def\J{{\rm J}}
\def\Jcoset{{\rm J}_{\rm coset}}
\def\K{{\rm K}}
\def\ident{{\equiv}}
\def\mimo{{\calc}}
\def\mimoorb{\calc^{\omega}}
\def\calao{\cala^{\omega}}
\def\Nto{{\mathfrak A}^{\omega}}
\def\Ntvir{{{\cal V}}^{2}}
\def\un{{}^{\rm un}}
\def\us{{}^{\rm us}}
\def\tw{{}^{\rm tw}}
\def\plus{{}_{{}^+}}
\def\minus{{}_{{}^-}}
\def\plusminus{{}_{{}^\pm}}
\def\gsq#1#2{%
    {\scriptstyle #1}\square\limits_{\scriptstyle #2}{\,}} % Ginsparg square
\def\sqr#1#2{{\vcenter{\vbox{\hrule height.#2pt  
 \hbox{\vrule width.#2pt height#1pt \kern#1pt
 \vrule width.#2pt}\hrule height.#2pt}}}}
\def\square{%
  \mathop{\mathchoice{\sqr{12}{15}}{\sqr{9}{12}}{\sqr{6.3}{9}}{\sqr{4.5}{9}}}}

\def\comment#1{\vskip0.2cm \br\vbox{\ni ||\  {\tt #1} \ ||}\vskip0.2cm}

\newcommand{\boubra}[1]  {{\langle\!\langle#1|}}
\newcommand{\bouket}[1]  {{|#1\rangle\!\rangle}}
\newcommand{\ishi}[1]    {{\langle\!\langle#1\|}}
\newcommand{\bashi}[1]   {{\|#1\rangle\!\rangle}}

%%%%%%%%% Christian's mymacros
%%%%%%%%%%%%

\def\yboxit#1#2{\vbox{\hrule height #1 \hbox{\vrule width #1
\vbox{#2}\vrule width #1 }\hrule height #1 }}
\def\fillbox#1{\hbox to #1{\vbox to #1{\vfil}\hfil}}
\def\ybox{{\lower 1.3pt \yboxit{0.4pt}{\fillbox{8pt}}\hskip-0.2pt}}
\def\mapr{\mathop{\longrightarrow}\limits}
\def\grade{\varphi}
\def\l{\left}
\def\r{\right}
\def\comments#1{}
\def\cc{{\rm c.c.}}
\def\tM{\tilde M}
\def\tN{\tilde N}
\def\bM{\bar M}
\def\QC{\Bbb{C}}
\def\QH{\Bbb{H}}
\def\QM{\Bbb{M}}
\def\QR{\Bbb{R}}
\def\QX{\Bbb{X}}
\def\QZ{\Bbb{Z}}
\def\p{\partial}
\def\tilp{\tilde\partial}
\def\eps{\epsilon}
\def\half{{\frac12}}
\def\pder#1#2{{\frac{\partial{#1}}{\partial{#2}}}}
\def\oder#1#2{{\frac{d{#1}}{d{#2}}}}
\def\cint#1#2{{\oint_{#1}\frac{d#2}{2\pi i}}}
\def\Tr{{{\rm Tr}}}
\def\tr{{\rm tr}}
\def\Re{{\rm Re\hskip0.1em}}
\def\Im{{\rm Im\hskip0.1em}}
\def\Ext{{\rm Ext}}
\def\adj{{\rm adj}}
\def\even{{\rm even}}
\def\odd{{\rm odd}}
\def\lcm{{\rm lcm}}
\def\diag{{\rm diag}}
\def\bra#1{{\langle}#1|}
\def\ket#1{|#1\rangle}
\def\bbra#1{{\langle\langle}#1|}
\def\kket#1{|#1\rangle\rangle}
\def\rrangle{\rangle\rangle}
\def\llangle{\langle\langle}
\def\vev#1{\langle{#1}\rangle}
\def\limitsvev#1{\left\langle{#1}\right\rangle}
\def\Dslash{\rlap{\hskip0.2em/}D}
\def\CA{{\cal A}}
\def\CC{{\cal C}}
\def\CD{{\cal D}}
\def\CE{{\cal E}}
\def\CF{{\cal F}}
\def\CG{{\cal G}}
\def\CT{{\cal T}}
\def\CL{{\cal L}}
\def\CM{{\cal M}}
\def\CN{{\cal N}}
\def\CO{{\cal O}}
\def\CP{{\cal P}}
\def\CQ{{\cal Q}}
\def\CS{{\cal S}}
\def\CV{{\cal V}}
\def\CW{{\cal W}}
\def\CX{{\cal X}}
\def\ad#1#2{{\delta\over\delta\sigma^{#1}(#2)}}
\def\ppt{{\partial\over\partial t}}
\def\comment#1{[#1]}
\def\nl{\hfill\break}
\def\ap{\alpha'}
\def\floor#1{{#1}}
\def\sgn{{\rm sgn\ }}
\def\P{\BP}
\def\I{I}
\def\IA{IA}
\def\II{\relax{I\kern-.10em I}}
\def\IIa{{\II}a}
\def\IIb{{\II}b}
\def\TeV{{\rm TeV}}
\def\hk{hyperk\"ahler\  }
\def\Hk{Hyperk\"ahler\  }
\def\cascade{{\cal A}}
\def\imp{$\Rightarrow$}
\def\IZ{\relax\ifmmode\mathchoice
{\hbox{\cmss Z\kern-.4em Z}}{\hbox{\cmss Z\kern-.4em Z}}
{\lower.9pt\hbox{\cmsss Z\kern-.4em Z}}
{\lower1.2pt\hbox{\cmsss Z\kern-.4em Z}}\else{\cmss Z\kern-.4em
Z}\fi}
\def\IB{\relax{\rm I\kern-.18em B}}
\def\IC{{\relax\hbox{$\inbar\kern-.3em{\rm C}$}}}
\def\ID{\relax{\rm I\kern-.18em D}}
\def\IE{\relax{\rm I\kern-.18em E}}
\def\IF{\relax{\rm I\kern-.18em F}}
\def\IG{\relax\hbox{$\inbar\kern-.3em{\rm G}$}}
\def\IGa{\relax\hbox{${\rm I}\kern-.18em\Gamma$}}
\def\IH{\relax{\rm I\kern-.18em H}}
\def\II{\relax{\rm I\kern-.18em I}}
\def\IK{\relax{\rm I\kern-.18em K}}
\def\IP{\relax{\rm I\kern-.18em P}}
%\def\IX{\relax{\rm X\kern-.01em X}}
%this doesn't work
\def\IX{{\bf X}}
\def\inbar{\,\vrule height1.5ex width.4pt depth0pt}
\def\mod{{\rm\; mod\;}}
\def\ndt{\noindent}
\def\p{\partial}
\def\pab{\pb_{\bar A} }
\def\pb{{\bar \p}}
\def\pgp{\pb g g^{-1}}
\font\cmss=cmss10 \font\cmsss=cmss10 at 7pt
\def\IR{\relax{\rm I\kern-.18em R}}
\def\pbar{\bar{\p}}
\def\qmvw{\CM_{\vec \zeta}(\vec v, \vec w) }
\def\sdtimes{\mathbin{\hbox{\hskip2pt\vrule
height 4.1pt depth -.3pt width .25pt\hskip-2pt$\times$}}}
\def\im{{\rm im\ }}
\def\ker{{\rm ker\ }}
\def\cok{{\rm cok\ }}
\def\End{{\rm End\ }}
\def\Hom{{\rm Hom}}
\def\id{{\it id}}
\def\rk{{r}}
\def\chch{\hbox{ch}}
\def\chcl{\hbox{c}}
\def\td{\hbox{Td}}
\def\aroof{\hat A}
\def\lroof{\hat L}
\def\euch{\hbox{e}}
\def\pocl{\hbox{p}}
\def\thocl{\Phi}
\def\BR{\IR}
\def\BZ{Z} % for now
\def\BP{\IP}
\def\BR{\IR}
\def\BC{\IC}
\def\BM{\QM}
\def\BH{\QH}
\def\BX{\QX}
\def\BN{N}
\def\Bid{{\mathchoice {\rm {1\mskip-4.5mu l}} {\rm
{1\mskip-4.5mu l}} {\rm {1\mskip-3.8mu l}} {\rm {1\mskip-4.3mu l}}}}
\def\ls{l_s}
\def\ms{m_s}
\def\gs{g_s}
\def\lp10{l_P^{10}}
\def\lp11{l_P^{11}}
\def\mb{{m_{\rm brane}}}
\def\vb{{v_{\rm brane}}}
%
%%%%%%%%%%%%

%%%%%%%%%%%%%%%%
%% Kris's macros
%%%%%%%%%%%%%%%%

\newcommand{\nc}{\newcommand}
\nc{\rnc}{\renewcommand}
\nc{\CY}{Calabi-Yau}
\nc{\CYM}{Calabi-Yau manifold}
\nc{\CYMs}{Calabi-Yau manifolds}
\nc{\DB}{D-Brane}
\nc{\DBs}{D-Branes}
\nc{\SUSY}{supersymmetry}
\nc{\Kah}{K\"ahler}
\nc{\cs}{complex structure}
\nc{\beq}{\begin{equation}}
\nc{\eeq}{\end{equation}}
\nc{\ntwo}{${\cal N}=2$}
\nc{\nOne}{${\cal N}=1$}
\nc{\hs}{\hspace{0.2in}}
\nc{\Z}{{\mathbb Z}}
\rnc{\P}{{\mathbb P}}
\rnc{\RP}{{\mathbb {RP}}}
%\nc{\R}{{\mathbb R}}
%\nc{\C}{{\mathbb C}}
\nc{\WP}{\mathbb{WP}}
\nc{\slag}{special Lagrangian}
\nc{\cn}{\C^n}
\nc{\rn}{\R^n}
%\nc{\M}{{\cal M}}
%\nc{\W}{{\cal W}}
\def\ket#1{|#1\rangle}
\def\kket#1{|#1\rangle\rangle}
%\nc{\SO}{\hbox{SO}}
%\nc{\Sp}{\hbox{Sp}}
%\nc{\SU}{\hbox{SU}}
\nc{\SO}{SO}
\def\Sp{Sp}
\nc{\SU}{SU}

\nc{\Wtree}{W_{\mathrm tree}}
\nc{\Weff}{W_{\mathrm eff}}

\def\vect#1#2#3{{ \left( \!\!
\begin{array}{c} #1 \\ #2 \\ #3 \end{array} \!\! \right)} }

\def\itsubsub#1{{\medskip\noindent\it #1\nopagebreak\medskip}}
%%%%%%%%%%%%

\catcode`\@=12

%\begin{document}

\pagestyle{myheadings}
\markboth{BRANDHUBER, ITA, NIEDER, OZ, AND R\"OMELSBERGER}{CHIRAL RINGS, SUPERPOTENTIALS AND ...}
\thispagestyle{empty}

\setcounter{page}{269}
\copyrightnotice{2003}{7}{269}{305}

\title{\Large \bf Chiral Rings, 
Superpotentials and the Vacuum Structure of 
${\cal N}=1$ Supersymmetric Gauge Theories}

\pubnum{%
CALT-68-2431 \\
USC-03-01 \\
TAUP-2722/03 \\
hep-th/0303001}

%\date{February 2003}
\date{}

\author{Andreas Brandhuber$^\flat$, Harald Ita$^\natural$, 
Harald Nieder$^\natural$, \\ Yaron Oz$^\natural$ 
and Christian~R\"omelsberger$^\dagger$ \\[0.4cm]
\rm  ${}^\flat$Department of Physics \\
\rm  California Institute of Technology\\
\rm  Pasadena, CA 91125, USA \\[0.2cm]
\rm  ${}^\natural$  Raymond and Beverly Sackler Faculty of Exact Sciences\\
\rm  School of Physics and Astronomy\\
\rm  Tel-Aviv University , Ramat-Aviv 69978, Israel\\[0.2cm]
\rm  ${}^\dagger$Department of Physics and Astronomy \\
\rm  University of Southern California \\
\rm  Los Angeles, CA 90089, USA \\[1.0cm]
}

\begin{abstract}We use the Konishi anomaly equations to construct the exact
effective superpotential of the glueball superfields in various
${\cal N}=1$ supersymmetric gauge theories. 
We use the superpotentials to study in detail the structure of the
spaces of vacua of these theories.
We consider chiral and
non-chiral $SU(N)$ models, the exceptional gauge group $G_2$ and
models that break supersymmetry dynamically.
\end{abstract}

%\enlargethispage{1.5cm}
%\makepapertitle
%\maketitle

%\vfill \eject 
\baselineskip=.16in
\tableofcontents

%\body

\version\versionno

\section{Introduction}

The strong coupling dynamics of ${\cal N}=1$ supersymmetric gauge 
theories in four dimensions is clearly of much 
theoretical and maybe physical importance.
Recently Dijkgraaf and Vafa made a beautiful conjecture 
\cite{Dijkgraaf:2002fc,Dijkgraaf:2002vw,Dijkgraaf:2002dh}
that the F-terms of a large
class of ${\cal N}=1$ supersymmetric gauge theories can be computed
exactly by a large $N$ computation in a bosonic matrix model.
The assumption is that the relevant fields in the IR are the glueball
superfields $S_i$ and the conjecture provides means of computing their 
exact effective superpotential.
This is done by evaluating the planar diagrams of the matrix model.
The generic glueball superpotential is a sum of 
Veneziano-Yankielowicz
logarithmic superpotential terms \cite{Veneziano:1982ah} 
and an infinite perturbative sum in the $S_i$.
Thus, even if the matrix model is not solvable, one can still compute
the superpotential to arbitrary power of $S_i$ by evaluating matrix
model diagrams.

In \cite{Cachazo:2002ry} it has been shown for a theory with adjoint matter 
that the loop equations for the matrix
model associated with the ${\cal N}=1$ gauge theory are equivalent to the generalized Konishi
anomaly equations. Besides being a nice observation by itself, the
loop equations can sometimes be powerful enough in order to solve 
the large $N$ matrix model. Also, one can forget about the 
matrix model and study the Konishi anomaly equations by themselves.
This will be the approach that we will take in this paper.

The aim of this paper is to study various types of  ${\cal N}=1$
supersymmetric gauge theories and to compute the exact glueball effective 
superpotential by using the Konishi anomaly.
%As we will see this method is very fast and powerful. 
We analyze the vacuum structure of those theories.
This approach works for chiral and non-chiral theories as well as theories
with exceptional gauge groups. One can even study theories with
dynamical supersymmetry breaking.

The paper is organized as follows: 
In section \ref{sec:chiralring} we discuss the general strategy and its 
limitations.
Limitations can be of different sorts. One could be that there are
not enough equations to solve for the superpotential.
Another requirement is the existence of a supersymmetric vacuum, which
limits the analysis of models that break supersymmetry.
In section \ref{sec:qcd} we analyze $SU(N)$ gauge theory with matter in the
fundamental representation and a quartic superpotential.
We analyze the different classical and quantum vacua and compute
the exact quantum superpotentials. We show how motions in the
parameter space of the theory interpolate between different vacua.
In section \ref{sec:g2} we analyze a gauge theory based on the exceptional
group $G_2$ with matter in the fundamental representation. Again, we
compute the exact superpotential and discuss the vacuum structure.
In section \ref{sec:ciralmodels} we discuss a chiral model and 
perform a similar analysis.
Results are in agreement with other methods.
In section \ref{sec:dsb} we analyze the IYIT model that breaks 
supersymmetry dynamically.
%Since our method works only in supersymmetric vacua, we will first deform
%the theory by a mass term and calculate the effective superpotential. We 
%will the study the behavior of the vacua, when the mass deformation
%is turned off. Requiring that we are left with a supersymmetric vacuum, will
%impose a constraint on the remaining coupling constants.
In the appendix we prove (under certain conditions)
the one-loop exactness of the (generalized) Konishi anomaly
equation. This result is used in the previous sections.

\section{The Classical and Quantum Chiral Rings}\label{sec:chiralring}

In this section we discuss some aspects of the classical and quantum
chiral rings of four-dimensional supersymmetric gauge theories.
The discussion parallels the one in \cite{Cachazo:2002ry}.
 
We denote the four-dimensional Weyl spinor supersymmetry generators by 
$Q_{\alpha}$ and $\bar{Q}_{\dot{\alpha}}$.
Chiral operators are operators annihilated by
$\bar{Q}_{\dot{\alpha}}$.
For instance, the lowest component $\phi$ of a chiral superfield
$\Phi$ is a chiral operator. 
The OPE of two chiral operators is nonsingular and allows for the definition 
of the product of two chiral operators. The product of chiral operators is 
also a chiral operator. 
Furthermore, one can define a ring structure on the set of 
equivalence classes of chiral operators modulo operators of the form
$\{\bar{Q}_{\dot{\alpha}}$,$\cdots \left.\right]$.

Consider a general $\mathcal{N}=1$ supersymmetric 
gauge theory with gauge group $G$ and some matter supermultiplets.
Denote by $V$ the vector superfield in the adjoint representation
of $G$, by $\Phi$ chiral superfields in a representation 
$r$ of $G$ and by $\phi$ their lowest component. 
The field strength (spinor) superfield is $W_\alpha =
-\frac{1}{4}\bar D^2e^{-V}D_\alpha e^V$ and  is a chiral superfield.
Using products of $\phi$ and $W_{\alpha}$ we construct chiral 
operators\footnote{In this paper we denote by $W_\alpha$ 
the supersymmetric field strength as well as its lowest component, 
the gaugino.}.
They satisfy the relation 
\beq
W_\alpha^{(r)}\phi^{(r)}=0
%~~~~mod~~~~\{\bar{Q}_{\dot{\alpha}},...\left.\right]
%\ 
\label{rel}
\end{equation}
modulo $\{\bar{Q}_{\dot{\alpha}}$,$\cdots \left.\right]$ terms, where we noted that $\phi$ transforms in a 
representation $r$ of the
gauge group $G$. $W_\alpha^{(r)}= W_{\alpha}^aT^a(r)$ with $T^a(r)$
being the generators of the gauge group $G$ in the representation $r$.
The relation (\ref{rel}) implies in particular
\beq
\{W_\alpha^{(r)},W_\beta^{(r)}\}=0 ~.
\end{equation}

We will be interested in the sector of gauge invariant chiral operators. 
These can be constructed as gauge invariant composites of 
$W_\alpha$ and $\phi$ taking the identity (\ref{rel}) into account.
We will call the chiral ring the ring of equivalence classes
of gauge invariant chiral operators modulo 
$\{\bar{Q}_{\dot{\alpha}},...\left.\right]$.
An important element of the chiral ring is the glueball superfield $S$
\beq
S=-\frac{1}{32\pi^2}\Tr W^2 ~.
\end{equation}

%into account\footnote{One might wonder, 
%why $W_\alpha$ is not by definition $\bar Q$-exact, since 
%$W_\alpha=-\frac{1}{4}\bar D^2e^{-V}D_\alpha e^V$. This is due to the
%fact that $\bar D_{\dot\alpha}e^{-V}D_\alpha e^V$ is not a globally
%well defined field. Much like the gauge field strength $F_{\mu\nu}$
%is not trivial.}. 

The gauge invariant chiral operators made of the matter multiplets
para\-me\-trize the moduli space of
vacua of the supersymmetric gauge theory.
It is therefore of interest to find the relations among the
elements of the chiral ring which
constrain the structure of the moduli space. These relations can be 
different in the classical and in the quantum theory. 
%
%The relation (\ref{chirrela})
%however is true in both cases, 
%since it is only used for the UV definition of the gauge invariant chiral 
%operators.

\subsection{The Classical Chiral Ring Relations}

Let us comment first on the relations in the classical chiral ring. 
There are two types of relations. The first are kinematic ones
which are associated with group theory and statistics and contain no 
dynamics of the classical theory. One such relation is
\beq
S^{\dim G+1}=0 ~,
\end{equation}
or an even stronger relation
\beq\label{chirrelb}
S^{h^\vee}=0 ~,
\end{equation}
where $h^\vee$ is the dual Coxeter number of the group $G$.
The last relation has been proven in \cite{Cachazo:2002ry} for
$G=SU(N)$ and in \cite{Witten:2003} for gauge groups $Sp(N)$ and $SO(N)$.
It has been  conjectured to hold for all simple groups.
Another example of a  kinematic relation is
\beq
\Tr\phi^n=\CP(\Tr\phi,\cdots,\Tr\phi^N) ~,
\end{equation}
with $\phi$ an adjoint field in a $U(N)$ gauge theory and $n>N$.

The second type of relations in the classical ring are the
dynamical relations given by the variation of the tree level 
superpotential $\Wtree$
\beq\label{chirrelc}
\pder{\Wtree}{\phi}=0 ~.
\end{equation}
These relations are not gauge invariant but can be implemented in a
gauge invariant way.
For instance by 
\beq
\phi\,\pder{\Wtree}{\phi}=0 ~,
\end{equation}
with appropriate extraction of the gauge invariant parts of the equations.

For a generic tree level superpotential the relations (\ref{chirrelc})
fix the moduli space of the theory up to a discrete choice. 
This means that we can solve these relations 
(possibly together with kinematic relations) 
and fix all the gauge invariant chiral operators made out of matter fields.

\subsection{The Quantum Chiral Ring Relations}

The classical chiral ring relations have quantum deformations. 
In general it is hard to find the quantum deformations unless 
there are enough symmetries in the theory.
However the classical relations arising 
from (\ref{chirrelc}) have a natural 
generalization as anomalous Ward identities of the quantized matter 
sector in a classical gauge(ino) background. 
If $\phi$ transforms in a representation 
$r$ of the gauge group $G$, 
then the classical superpotential relation (\ref{chirrelc}) 
transforms in the dual representation $\bar r$.
It has to be contracted with a chiral operator  $\phi'$ in a
representation $r'$ 
such that the decomposition of the
tensor product $\bar r\otimes r'$ contains a singlet
representation. This yields a classical chiral ring relation
\beq
\phi'\,\pder\Wtree\phi=0 ~.
\end{equation}
This relation can be interpreted as a classical Ward identity for the
Konishi current
$J=\Phi^\dagger e^V \Phi'$
\beq
\bar D^2J=\phi'\,\pder\Wtree\phi ~.
\end{equation}

This Ward identity gets an anomalous contribution in the quantum
theory. In general $\phi'$ is a function of $\phi$ and the generalized 
Konishi anomaly takes the form 
\cite{Cachazo:2002ry,Konishi:1984hf,Konishi:1985tu}
\beq\label{genkonishi}
\bar D^2J=
\phi'_i\pder\Wtree{\phi_i}+\frac{1}{32\pi^2}W_\alpha{}_i{}^j
W^\alpha{}_j{}^k\pder{\phi'_k}{\phi_i}~,
\end{equation}
where $i$, $j$ and $k$ are gauge indices and their contraction
is in the appropriate representation. 
This Ward identity has tree level and one loop contributions. 
In order to prove that there are no higher loop or nonperturbative 
corrections to this identity one has to use symmetry arguments 
and asymptotic behavior in the coupling constants 
(see the appendix for a proof under certain conditions).

%Since the divergence of $\bar D^2J$ is
%$\{\bar{Q}_{\dot{\alpha}},\cdots\left.\right]$ 
Since the divergence $\bar D^2J$ is $\bar{Q}$-exact it vanishes in a 
supersymmetric vacuum. Taking the Wilsonian expectation value of 
(\ref{genkonishi}) in a slowly varying gaugino background, we get
\beq\label{chirreld}
\limitsvev{\phi'_i\pder\Wtree{\phi_i}}_S
+\limitsvev{\frac{1}{32\pi^2}W_\alpha{}_i{}^j
W^\alpha{}_j{}^k\pder{\phi'_k}{\phi_i}}_S
= 0 ~.
\end{equation}
This relation will be our main tool to determine
the effective superpotential.
%We will see that this relation can be powerful enough 
%to determine the effective superpotential $\Weff$ of the glueball
%superfield $S$.

\subsection{The Effective Superpotential}

%We know less relations in the quantum chiral ring than 
%in the classical one. In particular, 
%we have no known quantum counterpart of (\ref{chirrelb}),
%unless the theory is pure glue. 
%This makes it hard to find the entire quantum chiral ring. 
We will be interested in determining the effective 
superpotential $\Weff$ for the glueball superfield 
$S$ with the matter fields $\Phi$ being integrated out. 

The strategy we will use is as follows.
We first use the gradient equations for $\Weff$ in the tree level 
superpotential couplings. For a tree level superpotential 
\beq
\Wtree=\sum_I g_I \sigma_I ~,
\end{equation}
where  $\sigma_I$ are gauge invariant chiral operators,  we get
\beq\label{gradeqn}
\pder\Weff{g_I}=\vev{\sigma_I}_S ~.
\end{equation}
The expectation values are taken in a slowly varying (classical)
gaugino background. 
We then use the chiral ring relations (\ref{chirreld}) to solve for the 
$\vev{\sigma_I}_S$ in terms of the $S$ and the coupling constants $g_I$. 
In order to solve these relations we use the factorization property
\beq
\vev{\sigma_I\sigma_J}_S=\vev{\sigma_I}_S\vev{\sigma_J}_S 
\end{equation}
of expectation values of chiral operators in a supersymmetric vacuum.

We insert the solutions into the gradient equations (\ref{gradeqn}) 
and determine the effective superpotential up to a function $C(S)$, 
which does not depend on the $g_I$. 
%This function should typically be of a Veneziano-Yankielowicz form. 
We can determine this function by 
semi classical arguments in certain limits of the coupling constant space,
%where all matter is massive. 
 where the low energy dynamics is described by pure SYM.
%In such a limit the strong IR dynamics 
%is captured by a Veneziano-Yankielowicz type superpotential.
The strong IR gauge dynamics is then captured by a Veneziano-Yankielowicz 
type superpotential.
If there are several such limit points, there are consistency checks one 
can do. 
%This method fails when supersymmetry is broken!

%Again, for a generic tree level superpotential, 
%one can solve the relations (\ref{chirreld}) for all the chiral 
%singlet fields, which contain matter fields only. 
%But now the solution will depend on other singlet fields, 
%which contain gauginos. 
%In general, this might be a problem, 
%if there are other fields involved, apart from the gaugino bilinear
%$S$. 
%Also, if there are group theoretic relations involved in the solution,

%one has to find the quantum deformations thereof 
%beforehand, e.g. In the case of $SU(N)$ with $N$ 
%flavors one needs to find the quantum deformation of the relation
%\beq
%\tilde B B=\det M
%\eeq
%between the mesons and the baryons. These relations might be of
%nonperturbative 
%nature and very hard to handle unless there is a lot of symmetry in
%the problem. 
%In the following section, we will go through several examples, 
%to show how this method can be applied and also how it can fail.

\section{$SU(N_c)$ with Fundamental Matter}\label{sec:qcd}

In this section we will consider ${\cal N}=1$ supersymmetric gauge
theories with $SU(N_c)$ gauge group and matter in the 
fundamental representation. We will use SQCD with one flavor
as a representative model to outline our technique.
Due to the small number of generators in the chiral ring
the usual Konishi anomaly suffices to obtain the effective superpotential.
Using the full effective superpotential we will
then analyze the vacuum structure of the model.
We discuss also various generalizations.   

\subsection{$SU(N_c)$ with One Flavor}

Let us start by considering SQCD with gauge group $SU(N_c)$ and one flavor.
This theory was studied in the matrix model context in 
\nocite{Seiberg:2002jq,Argurio:2002xv,McGreevy:2002yg,Suzuki:2002gp,
Bena:2002kw,
Demasure:2002sc,
Tachikawa:2002wk,Argurio:2002hk,Naculich:2002hr,Bena:2002ua,Feng:2002zb,
Feng:2002yf,Ookouchi:2002be,Ohta:2002rd,Bena:2002tn,Hofman:2002bi,
Roiban:2003uq,Feng:2002is,Demasure:2002jb,Suzuki:2002jc}
\cite{Seiberg:2002jq}-\cite{Suzuki:2002jc}.
The theory is 
non-chiral and has one fundamental matter multiplet $Q$ and one 
antifundamental matter multiplet $\tilde Q$. There are only two gauge 
invariant chiral operators one can build out of the fundamental
fields, the meson $M=\tilde QQ$ and the 
gaugino bilinear $S$. To begin with let us assume the tree level 
superpotential
\begin{equation}\label{wtree}
\Wtree=mM + \lambda M^2 ~.
\end{equation}
Note we have chosen a rather simple superpotential to illustrate our
method but in principle we could take an arbitrary polynomial in the
meson field. The theory with (\ref{wtree})
has two classical vacua at $M=0$ and $M=-\frac{m}{2\lambda}$. 
In the first vacuum the gauge group is unbroken, whereas in the second 
vacuum the gauge group is broken to $SU(N_c-1)$.

The Konishi variation $\delta Q=\epsilon Q$ leads to the
relation
\begin{equation}
m\vev{M}_S + 2 \lambda \vev{M^2}_S =S ~.
\end{equation}
%which is like Seiberg`s loop equation \cite{Seiberg:2002jq}
%\begin{equation}
%\left[M(z)m(z)\right]=R(z)
%\end{equation}
%at $O(\frac{1}{z})$. 
We also have the relations
\begin{equation}
\frac{\partial \Weff}{\partial m}= \vev{M}_S ~, \quad 
\frac{\partial \Weff}{\partial \lambda}= \vev{M^2}_S ~.
\end{equation}
If we use the factorization properties for the chiral operators 
we get a quadratic equation for $\vev{M}_S$
\begin{equation}\label{qcdkonishi}
2 \lambda \vev{M}_S^2 +m \vev{M}_S -S=0 ~,
\end{equation}
Now we get two differential equations which control the 
dependence of the effective superpotential on the bare couplings 
\begin{equation}
\label{eq:qcdkonishi2}
\frac{\partial \Weff}{\partial m}=-\frac{m}{4\lambda}  \pm 
\sqrt{\frac{m^2}{16 \lambda^2}+\frac{S}{2\lambda}} ~,
\end{equation}
\begin{equation}
\label{eq:qcdkonishi3}
\frac{\partial \Weff}{\partial \lambda}=\left(-\frac{m}{4\lambda} \pm 
\sqrt{\frac{m^2}{16 \lambda^2}+\frac{S}{2\lambda}}\right)^2 ~.
\end{equation}
By taking the classical limit $S \to 0$ in the above equations we find that
the $+$ sign corresponds to the classical vacuum $M=0$ and the $-$ sign
to $M=-\frac{m}{2 \lambda}$.
These two equations can be integrated to give the following superpotential
\begin{equation}
\label{eq:Weff1}
\Weff=-\frac{m^2}{8\lambda} \pm \frac{m^2}{8\lambda}\sqrt{1+\frac{8 \lambda}
{m^2}S}+S\log{m}+S\log{(1 \pm \sqrt{1+\frac{8 \lambda}{m^2}S})}+C(S) ~,
\end{equation}
where $C(S)$ is an $S$ dependent integration constant.

%new stuff
 To determine $C(S)$ we proceed as follows. In the classical vacuum $M=0$
the matter fields have mass $m$. If we take $m >> \Lambda$, where 
$\Lambda$ is the dynamically generated mass scale we can integrate out
the matter fields in perturbation theory. 
Hence, we separate the pure gauge dynamics from the dynamics of the
matter fields. We will take care of the pure gauge dynamics in the strong
coupling regime by an appropriate Veneziano-Yankielowicz term \cite{Veneziano:1982ah}.
We concentrate first on an 
effective action $W_{eff}^{pert.}(S)$ for $S$
obtained by integrating out the matter fields in perturbation theory.
As explained in \cite{Cachazo:2002ry} the terms of $W_{eff}^{pert.}(S)$ 
linear in $S$ come from integrating out the matter fields at one loop.
Higher loops depend on the bare couplings in the tree level superpotential
and are thus already included in (\ref{eq:Weff1}) \cite{Dijkgraaf:2002xd}.
The contribution $C(S)$ can thus be determined by an explicit one loop calculation \cite{Cachazo:2002ry}.

Note that in general this requires that we have a classical vacuum where all the 
matter fields are massive around which we can do perturbation theory.
However, this method also works for vacua
where all the matter degrees of freedom are eaten up by the Higgs mechanism.
This will prove especially useful in the case of chiral models
where we cannot have mass terms for the matter fields.
%old stuff

%The contribution $C(S)$ can be determined by demanding that the above 
%superpotential reproduces the perturbative result. 
Right now we consider perturbation theory around the classical vacuum $M=0$.
%The appropriate branch in the superpotential (\ref{eq:Weff1}) thus corresponds to the
%$+$ sign.
The perturbative superpotential at an energy scale $\Lambda<\mu < m$ is given by
\begin{equation}
\label{eq:Wpert}
\Weff^{pert.}=\tau_0 S + 3N_cS\log{\frac{\Lambda_0}{\mu}}+
S\log{\frac{m}{\Lambda_0}}+O(S^2) ~,
\end{equation}
where $\Lambda_0$ is the UV cutoff. Substituting
\begin{equation}
\tau_0=-(3N_c-1) \log{\frac{\Lambda_0}{\Lambda}}
\end{equation}
amounts to replacing $\Lambda_0$ by $\Lambda$ in (\ref{eq:Wpert}).
Since we compare (\ref{eq:Weff1}) to a one loop calculation around the vacuum $M=0$
we have to choose the branch with the $+$ sign.
Matching then the contributions of $O(S)$ in (\ref{eq:Weff1}) and 
(\ref{eq:Wpert}) gives
\begin{equation}
C(S)=3N_cS\log{\frac{\Lambda}{\mu}}-S\log{\Lambda}-\frac{S}{2}-S\log{2} ~.
\end{equation}
We have to include the strong coupling dynamics by replacing
\footnote{This amounts to replacing $\mu^3 \to S/e$.} 
\begin{equation}
3N_cS\log{\frac{\Lambda}{\mu}} \to N_cS\left(-\log{\frac{S}{\Lambda^3}}+1 
\right) ~.
\end{equation}
%{\it This corresponds to changing $\mu^3 \to S/e$.}
Finally matching the scale $\hat{\Lambda}$ of the pure SYM according to
\begin{equation}
\hat{\Lambda}^{3N_c}=\Lambda^{3N_c-1}m ~,
\end{equation}
we find the full nonperturbative superpotential
\begin{eqnarray}
\label{eq:Weff2}
 \Weff & = & N_cS\left(-\log{\frac{S}{\hat{\Lambda}^3}}+1 \right) 
-\frac{S}{2}-\frac{m^2}{8\lambda}
 \pm \frac{m^2}{8\lambda}\sqrt{1+\frac{8\lambda}{m^2}S}+ \\ \nonumber
& & +S\log{\left(\frac{1}{2} \pm \frac{1}{2}
\sqrt{1+\frac{8\lambda}{m^2}S}\right)} ~. 
\end{eqnarray}
%{\it For future reference 
%we point out that from a technical point of view it proves more convenient to 
%include the strong coupling dynamics from the outset.
%Again we require 
%the theory to be described below a certain scale by pure SYM up
%to the corrections induced by the tree level superpotential.
%Then we can plug the Veneziano-Yankielowicz part for this pure SYM which accounts
%for the terms below $O(S^2)$ (there are also $S \log{S}$ terms) in the full
%effective superpotential. As before, the remaining higher order terms $O(S^2)$
%are controlled by the integrated Konishi relations.}

\itsubsub{Alternative Derivation of $W_{eff}$}

To gain confidence in this result, we can give an alternative 
derivation of (\ref{eq:Weff2}) based upon the 
ILS linearity principle \cite{Intriligator:1994jr}. 
If we consider the tree level superpotential as a perturbation to the low energy physics
we can first forget about the superpotential and consider $SU(N_c)$ SYM with one massless
flavor. Along the flat direction parametrized by $M$ the gauge group is generically broken
to $SU(N_c-1)$.
%Depending on the tree level superpotential 
%we can have several classical meson expectation values. For a vanishing vev 
%the gauge group is classically unbroken. In the quantum theory, however,
%the dynamically generated superpotential will give
%rise to a generically non-vanishing expectation value 
%and the gauge group is Higgsed.
%This is also obvious from the Konishi relation (\ref{qcdkonishi}).
We thus expect an appropriate effective description in terms of a pure $SU(N_c-1)$
SYM obtained by Higgsing the original $SU(N_c)$ with one flavor.
The effective superpotential for the $SU(N_c-1)$ theory is just
the Veneziano-Yankielowicz superpotential
\begin{equation}
\label{HVY1}
W_{eff}=\left(N_c-1\right) S \left[ -\log{\frac{S}{\tilde{\Lambda}^3}}+1 
\right] ~.
\end{equation}
We have to relate the scale $\tilde{\Lambda}$ of the Higgsed theory 
to the scale $\Lambda$ of the original theory. This is 
done at the scale set by the meson expectation value $M$. We have
\begin{equation}
\left(\frac{\tilde{\Lambda}}{M^{1/2}}\right)^{3(N_c-1)}=
\left(\frac{\Lambda}{M^{1/2}}\right)^{3N_c-1} ~,
\end{equation}
such that
\begin{equation}
\tilde{\Lambda}^{3(N_c-1)}=\frac{\Lambda^{3N_c-1}}{M} ~.
\end{equation}
Adding the tree level potential will localize the meson expectation value
at the quantum vacuum. This localization is equivalent to integrating out
$M$ from the effective superpotential. However, the quantum expectation value of the
meson as a function of $S$ is given by the Konishi relation (\ref{eq:qcdkonishi2}).
So if we add the tree level superpotential (\ref{wtree}) to (\ref{HVY1})
and replace $M$ by the quantum expectation value $\vev{M}_S$ 
given by the Konishi relation as
\begin{equation}
\label{vev1}
\vev{M}_S=-\frac{m}{4 \lambda}\pm \sqrt{\frac{m^2}{16 \lambda^2}+
\frac{S}{2 \lambda}} ~,
\end{equation}
we exactly reproduce the full nonperturbative superpotential
given in (\ref{eq:Weff2}). This gives us a nice consistency check.

\itsubsub{Relation to the Vector Model}

The anomaly equation (\ref{qcdkonishi}) can also be derived from the zero 
dimensional vector model
\beq
\int dQ d\tilde Q e^{-\frac{1}{g_s}\Wtree(\tilde Q Q)} ~.
\end{equation}
The Ward identity for the variation $Q\mapsto Q+\epsilon Q$ is
\beq
g_s N_{VM}=\limitsvev{\pder{\Wtree}{Q}Q}_{VM}=m\vev{M}_{VM}+2
\lambda\vev{M^2}_{VM} ~.
\end{equation}
Making the identification $S=g_s N_{VM}$, one reproduces the anomaly equation. 
In the planar limit $\vev{M^2}_{VM}$ factorizes and we get the same result as 
before. 

\subsection{The Vacuum Structure}

The expression for the effective superpotential has two branches for the two 
signs of the square root. These correspond to the two classical vacua. One 
can make an expansion for small $\frac{8\lambda}{m^2}S$ in both branches to 
recover $N_c$ vacua in the one branch and $N_c-1$ vacua in the other. This is 
the expected result from the semiclassical analysis since we expect the 
unbroken gauge symmetries to confine in the IR.

The quantum vacua are at the critical points of $\Weff$, {\it i.e.} they satisfy
\beq\label{eq:qveq}
\log{\left[ \frac{\hat{\Lambda}^{3N_c}}{S^{N_c}}  
\left(\frac{1}{2} \pm \frac{1}{2}
\sqrt{1+\frac{8\lambda}{m^2}S}\right)\right]} = 0 ~.   
\end{equation}
This can be simplified to
\beq\label{eq:onflmod}
  \hat S^{2N_c-1}-\hat S^{N_c-1}-z=0~,
\quad \hat S = \frac{S}{\hat\Lambda^3}~,
\quad z=\hat\Lambda^3 \frac{2\lambda}{m^2}~.
\end{equation}
From this equation it is possible to understand the quantum parameter space. 
It is given as the complex surface, associated to the analytic continuation 
in $z$ of the gaugino condensate 
$S = \hat\Lambda^3 \hat S(z)$.

Naturally, from the polynomial equation we expect $2N_c-1$ sheets. 
There is an order $2N_c-1$ branching point at $z=\infty$ and 
an order $N_c-1$ branching point at $z=0$.  
The remaining $N_c$ branching points are double points located at
the roots of the equation
\begin{equation}\label{sqcd:roots}
z^{N_c} = (-)^{N_c} 
\frac{(N_c-1)^{N_c-1} {N_c}^{N_c}}{(2 N_c - 1)^{2 N_c-1}}~.
\end{equation}
At the double points two zeros of the polynomial (\ref{eq:onflmod}) 
coincide, such that its first derivative vanishes. 
Since this is equivalent to the second derivative 
of the superpotential, we generally expect massless glueballs 
at these points. This implies that the mass gap disappears at those
points, unless the K\"ahler potential also gets singular.

\begin{figure}[htb]
\epsfxsize=2.5in
\bigskip
\centerline{\epsffile{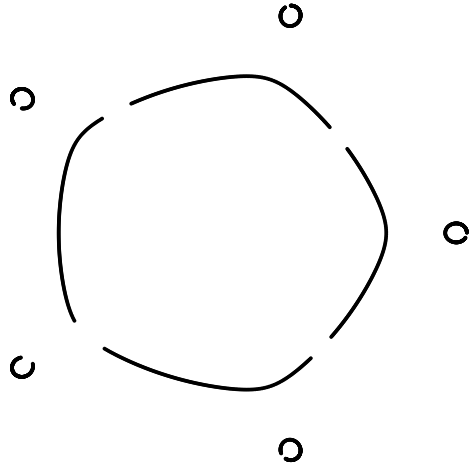}}
\caption{\label{fighiggs}
\footnotesize
Monodromy around $z=0$ for $G=SU(5)$: we follow the positions of the vacua in
the $S$ plane as we take a small, not completely closed circle
around $z=0$. 
The un-Higgsed vacua, corresponding to the smaller outer circles,
are not interchanged, whereas the big circle corresponds to
the four Higgsed vacua which get interchanged as we change the
phase of $z$.
}
\end{figure}

At the point $z=0$ we expect to find a clear distinction 
of Higgsed and un-Higgsed quantum vacua. And actually for small z the above 
equation factorizes to $S^{N_c-1}-\tilde\Lambda^{3(N_c-1)}=0$ 
(where we used $z\hat\Lambda^{3(N_c-1)}=\tilde\Lambda^{3(N_c-1)}$) 
and $S^{N_c}-\hat\Lambda^{3N_c}=0$. These give $N_c-1$ vacua for the 
Higgsed branch and $N_c$ for the un-Higgsed one with the appropriate scales 
$\tilde\Lambda$ and $\hat\Lambda$.

\begin{figure}[htb]
\epsfxsize=2.5in
\bigskip
\centerline{\epsffile{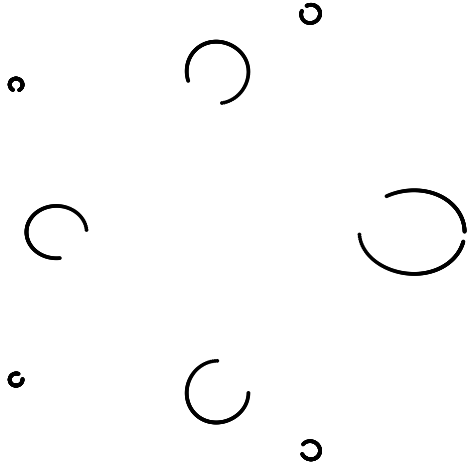}}
\caption{\label{figexch}
\footnotesize
Here we can see the motion of the vacua of the $SU(5)$ theory
in the $S$ plane as we vary the parameter $z$ along a small 
circle around a root of (\ref{sqcd:roots}). 
The un-Higgsed vacua correspond to the smaller outer circles,
whereas the bigger, inner circles correspond to the Higgsed vacua.
Most interesting is the rightmost big circle in which we can see
the exchange of an un-Higgsed with a Higgsed vacuum.
}
\end{figure}

Circling $z=0$ corresponds to rotating the scale $\tilde\Lambda^{3(N_c-1)}$ 
and thus changes from one quantum vacuum to the next in the Higgsed branch 
(see Fig. \ref{fighiggs}). 
Circling the bulk branch points 
changes from quantum vacua in the Higgsed branch to the un-Higgsed one 
(see Fig. \ref{figexch}). 
If we take $z$ to be large, then the $2 N_c - 1$ vacua arrange themselves
symmetrically on a circle and the monodromy at infinity $z \to z e^{2 \pi i}$
exchanges them in a $Z_{2 N_c - 1}$ symmetric manner.
So at $z=\infty$ both branches look similar, which seems natural, 
as the minima of $\Wtree$ degenerate ($z\rightarrow\infty$ is like 
$m\rightarrow 0$).  
The structure of the quantum parameter space for other models 
%and the way to move between the quantum vacua there 
has been discussed recently
in \cite{Ferrari:2002kq,Cachazo:2002zk,Ferrari:2003yr}.

\itsubsub{The Massless Limit}

It is interesting to analyze the massless limit of our SQCD model. 
As stated above this corresponds to the limit $z \to \infty$ in the
quantum parameter space.
From the 
analysis of the quantum parameter space we expect that the 
two branches join to give the $2 N_c-1$ vacua. 
Indeed, the effective superpotential (\ref{eq:Weff2}) has a finite 
$m \to 0$ limit on both branches
and we can recover the $2N_c-1$ vacua from either branch.
If we start on the un-Higgsed branch (the $+$ branch of (\ref{eq:Weff2}))
and take the massless limit we obtain
\begin{equation}
\label{Massl1}
W_{eff}=S \log{\Lambda^{3N_c-1}} +\left(N_c-\frac{1}{2} \right)S 
-\left(N_c-\frac{1}{2} \right)S \log{S} + \frac{1}{2}S
\log{\lambda}+\frac{1}{2}S \log{2} ~.
\end{equation}
Minimizing with respect to $S$ gives the $2N_c-1$ vacua
\begin{equation}
\label{ML1}
S=e^{\frac{4 \pi i k}{2N_c-1}}\,\left(2 \lambda\,\Lambda^{6N_c-2}\right)^{\frac{1}{2N_c-1}}~.
\end{equation}
We can compare this result with the Affleck-Dine-Seiberg analysis of the same
system. In this approach the exact effective superpotential\footnote{
The ADS superpotential can also be obtained from our 
glueball superpotential by a Legendre transform in $m$ and subsequent
integrating out of $S$.}
in the massless case is given by
\begin{equation}
W_{eff}^{ADS}=\left(N_c-1 \right)\left(\frac{\Lambda^{3N_c-1}}{M} 
\right)^{\frac{1}{N_c-1}} + \lambda M^2 ~.
\end{equation}
Looking for the mesonic vacua we find
\begin{equation}
\label{ML2}
M=\left(\frac{\Lambda^{3N_c-1}}{2 \lambda} \right)^{\frac{1}{2N_c-1}} 
e^{\frac{2 \pi i k}{2N_c-1}} ~.
\end{equation}
As expected the vevs (\ref{ML1}) and (\ref{ML2}) satisfy the Konishi relation
\begin{equation}
\label{MLK1}
2 \lambda \vev{M}_S^2  - S = 0 ~,
\end{equation}
obtained from (\ref{qcdkonishi}) in the massless limit. We thus get a nice 
picture of the $2N_c-1$ vacua of the massless model if we think of them as 
obtained in the massless limit of a massive model.

\subsection{$SU(N_c)$ with $N_c\ge N_f>1$}

We will now turn to a generalization of the above procedure to the case of
$N_c\ge N_f>1$, with a simple tree level superpotential. The effective superpotential
can be derived again with the use of the Konishi relations. With the 
superpotential at hand, we investigate the vacuum structure 
of the model. Many of the classical vacua 
turn out to be connected in the quantum parameter space. 
  
The model we consider is $SU(N)$ SQCD with $N_f$ flavors and tree 
level superpotential,
\begin{eqnarray}
  \label{eq:spmf}
  \Wtree=m\;\tr M+\lambda\; \tr M^2 ~,
\end{eqnarray}
with $M$ the meson $M_I{}^J=\tilde{Q}_I^iQ^J_i$. This superpotential breaks 
the $U(N_f)\times U(N_f)$ flavor symmetry to a diagonal $SU(N_f)$. The diagonally 
embedded $U(1)_B$ is responsible for the baryon number conservation.

The classical vacua can be understood in terms of the meson field $M$. 
First, we rotate the meson matrix $M_I^J$ to diagonal form by flavor 
rotations.
%the remaining $U(N_f)$. 
Then the tree level superpotential allows that $N_f^+$ eigenvalues
sit at zero and $N_f^-$ at the minimum $M_I^I=-m/2\lambda$. (With the condition
$N_f^++N_f^-=N_f$.) In total we have $N_f$ choices to distribute 
the eigenvalues of the meson.

Starting from a classical vacuum $(N_f^+,N_f^-)$ we have a clear expectation
 of the quantum theory for the parameter in the range 
$m/\lambda\gg\Lambda^2$ and $m\gg\Lambda$.
That is, for energies much higher than the meson expectation values $M_I^I=-m/2\lambda$ and 
the squark masses $m$, we expect to see $SU(N_c)$ gauge dynamics with $N_f$ 
almost massless quarks. (We assume an appropriate UV-completion of the above 
tree level potential.) Lowering the scale below the squark masses 
and the meson expectation values, but still 
above $\Lambda$, we expect to find pure $SU(N_c-N_f^-)$ supersymmetric gauge dynamics
with scale $\tilde\Lambda^{3(N_c-N_f^-)}=\Lambda^{3N_c-N_f}m^{N_f^+}(2\lambda/m)^{N_f^-}$.
Finally, for energies below $\tilde\Lambda$ one finds confinement 
with $N_c-N_f^-$ supersymmetric vacua. Starting from this well known vacuum 
structure we will extend the knowledge of the vacuum structure
to the case $m/\lambda<\Lambda^2$ and $m<\Lambda$ 
in the following.

To this end we have to use the Konishi relations. 
The flavor dependent Konishi anomaly 
variation $\delta Q^I_i=\lambda^I{}_J Q^J_i$ leads to
\begin{eqnarray}
  \label{eq:fdka}
 m\;M_I{}^J+2\lambda (M^2)_I{}^J=\delta_I^J  \;S ~.
\end{eqnarray}
We can solve for the diagonal entries of the meson $M$. 
Here we pick $N_f^+$ eigenvalues to converge to the vacuum $M=0$ and 
$N_f^-$ eigenvalues to converge to $M=-m/2\lambda$ in the classical limit, $S\rightarrow 0$. This amounts to 
choosing branches for each of the eigenvalues in (\ref{eq:fdka}). The traces 
then have the vacuum expectation values
\begin{eqnarray}
\vev{\tr M}_S&=&
-N_f^+\frac{m}{2\lambda}\left(\half-\half\sqrt{1+\frac{8\lambda}{m^2}S}\right)
-N_f^-\frac{m}{2\lambda}\left(\half+\half\sqrt{1+\frac{8\lambda}{m^2}S}\right) 
\nonumber\\
\vev{\tr M^2}_S&=&
N_f^+\frac{m^2}{4\lambda^2}\left(\half-\half
\sqrt{1+\frac{8\lambda}{m^2}S}\right)^2
+N_f^-\frac{m^2}{4\lambda^2}\left(\half+\half
\sqrt{1+\frac{8\lambda}{m^2}S}\right)^2. \nonumber
\end{eqnarray}
Note that this will break the diagonal $SU(N_f)$ flavor symmetry to $SU(N_f^+)\times 
SU(N_f^-)\times U(1)$, leaving $2N_f^+N_f^-$ massless Goldstone bosons.

In order to find the effective superpotential we can integrate 
the two gradient equations
\beq
\pder{\Weff}{m}=\vev{\tr M}_S, \;\;\;
\pder{\Weff}{\lambda}=\vev{\tr M^2}_S.
\end{equation}
By matching the integration constant, such that the appropriate VY potential is 
reproduced in the limit $m/\lambda\gg\Lambda^2$ 
and $m\gg\Lambda$, we can determine the effective superpotential 
\begin{eqnarray}
\label{eq:efsNf}
\Weff & = & N_cS\left(-\log{\frac{S}{\hat{\Lambda}^3}}+1 \right) 
-N_f(\frac{S}{2}+\frac{m^2}{8\lambda})
+(N_f^+-N_f^-)\frac{m^2}{8\lambda}\sqrt{1+\frac{8\lambda}{m^2}S}+ \nonumber\\
& & +S\log\left[{\left(\frac{1}{2} +\frac{1}{2}
\sqrt{1+\frac{8\lambda}{m^2}S}\right)^{N_f^+}\left(\frac{1}{2} -\frac{1}{2}
\sqrt{1+\frac{8\lambda}{m^2}S}\right)^{N_f^-}}\right] ~.   
\end{eqnarray}
This effective superpotential does not depend on the massless Goldstone 
modes, since they are true moduli by symmetry. This is like integrating 
out the radial direction in a Mexican hat potential.

The scales in this model are the UV scale $\Lambda$, the scale 
$\hat\Lambda$ for the theory with $N_f$ massive quarks around 
$Q=0$, and the scale $\tilde\Lambda$ for the theory with $N_f^+$ 
massive quarks around $Q=0$ and $N_f^-$ Higgsing quarks,
\begin{eqnarray}
   \label{eq:mhq}
   \hat\Lambda^{3N_c}=\Lambda^{3N_c-N_f}m^{N_f}=
 \tilde\Lambda^{3(N_c-N_f^-)}(m^2/2\lambda)^{N_f^-} ~.
\end{eqnarray}

\itsubsub{The Vacuum Structure}

Let us understand this result better in the limit of small $S$. One finds 
$N_c-N_f^-$ vacua with the appropriate scale plugged in. By analytic
continuation in the parameter space,
we can change the sign of the square roots, {\it i.e.} exchange the role of 
$N_f^+$ and $N_f^-$. Then for small $S$ we find $N_c-N_f^+$ vacua corresponding to $N_f^+$ 
Higgsing squarks. As in the case $N_f=1$,  these classical 
vacua are smoothly connected in the quantum parameter space.

The critical points of $\Weff$ are the supersymmetric ground states of
the theory. For a given branch of $\Weff$  
they are given by the following equation for $S$,
\beq\label{Nfbigger}
  \log{\left[ 
\frac{\hat{\Lambda}^{3N_c}}{S^{N_c}}  \left(\frac{1}{2} +\frac{1}{2}
 \sqrt{1+\frac{8\lambda}{m^2}S}\right)^{N_f^+}\left(\frac{1}{2} -\frac{1}{2}
 \sqrt{1+\frac{8\lambda}{m^2}S}\right)^{N_f^-}\right]} = 0 \ .   
\end{equation}
For $N_f < N_c$ this has $N_c - N_f^-$ solutions. However, we have to
take into account that the $+$ and $-$ branches can be distributed in
$\left(\begin{array}{c}\!\! N_f \\ r \end{array}\!\! \right)$ 
different ways on the $N_f$ meson eigenvalues. 
Hence, the total number of vacua is
\beq
\sum_{r=0}^{N_f} (N_c-r) \left( \!\! \begin{array}{c}
N_f \\ r \end{array} \!\! \right) = (2 N_c - N_f)\ 2^{N_f-1} ~.
\end{equation}

\itsubsub{The Case $N_f=N_c$}

 We would like to discuss now the above results
for the special case $N_f=N_c$. More specifically we consider the two branches:
$N_f^-=0$ and $N_f^- = N_c$.
In the first branch all meson vevs are zero
at the classical level, the gauge group is unbroken and confines in the IR 
giving rise to $N_c$ vacua. In the second case all meson vevs are non-zero, 
the gauge group is broken to nothing and classically there is a unique 
ground state.

Now we look for the quantum vacua by analyzing (\ref{Nfbigger}). It turns
out that the $N_f^-=0$ case has $N_c$ solutions and the vevs are given by
\begin{eqnarray}
  \label{eq:NceqNf}
  S&=&e^{2\pi i\,k/N_c}\hat\Lambda^3\left(1+e^{2\pi i\,k/N_c}
\frac{2\lambda\,\Lambda^2}{m}\right)\ , \nonumber \\
  \vev{\tr M}_S&=&e^{2\pi i\,k/N_c}\,N_f\,\Lambda^2 ~,
\end{eqnarray}
with $\Lambda^2m=\hat\Lambda^3$.
In the limit of small $\frac{2\lambda\,\Lambda^2}{m}$ the gaugino 
condensate reduces to $S=e^{2\pi i\,k/N_c}\hat\Lambda^3$, the vacua of 
pure $SU(N_c)$ gauge dynamics. There are $N_c$ points 
$m/2\lambda=-e^{2\pi i\,k/N_c}\Lambda^2$, where the gluino condensate vanishes.

The other case with $N_f^- = N_c$ is more subtle but a careful analysis of
(\ref{Nfbigger}) shows that there is an extremum at $S=0$ under the 
condition that 
$\left(-\frac{2 \lambda \hat\Lambda^3}{m^2}\right)^{N_c} = 1$. These are no 
new solutions, but just the points with $S=0$ from (\ref{eq:NceqNf}). 
The fact that the gauge group is completely broken on the $N_f^- = N_c$ branch
is consistent with the vanishing of the gluino condensate.
%{\it From the expectation values of the meson  Higgsed vacua exist classically for all values of 
%the parameters, in the quantum theory we find them only for specific points 
%in the parameter space.}

We see that the full parameter space has $N_c$ sheets, where 
each of them has two 
distinguished points, one corresponding to a vacuum of pure $SU(N_c)$ 
gauge dynamics and the other to a fully Higgsed vacuum. We will 
find this structure useful when considering dynamical SUSY breaking.

We would like to close this section with the observation, that the 
expectation value of the meson $M$ satisfies\footnote{Although we have discussed 
only the extremal cases $N_f^-=0$ and $N_f^-=N_c$ the quantum
constraint from (\ref{Nfbigger}) can be verified also for a generic eigenvalue 
distribution for $N_f=N_f^++N_f^-=N_c$.}
\begin{eqnarray}
  \label{eq:quntmodspa}
  Det\,M =\Lambda^{2N_c} ~, 
\end{eqnarray} 
all over the parameter space. On the other hand we can consider the 
generalized Konishi variation 
$\delta Q^I_i=\epsilon_{ij_2\ldots j_{N_c}}\epsilon^{IJ_2\ldots J_{N_c}}
\tilde Q^{j_2}_{J_2}\ldots\tilde Q^{j_{N_c}}_{J_{N_c}}$ which leads to
the relation
\beq
(mN_c+2\lambda \tr M)B=0,
\end{equation}
where $B=\det Q$ is the baryon. This implies $B=0$. Similarly, we can show
that $\tilde B=0$. This shows the validity of the relation
\beq
\det M-B\tilde B=\Lambda^{2N_c}~.
\end{equation}

\subsection{More General $\Wtree$}
% and the Integrability of the Gradient Equations}
\label{sec:gradient}

Let us now illustrate how to implement more general tree level 
superpotentials
\beq
\Wtree(M)=\sum_{j=1}^n\frac{g_j}{j}M^j ~.
\end{equation}
in SQCD with $N_f=1$. In that case the Konishi constraint yields
\beq
\sum_{j=1}^ng_j\vev{M}_S^j=S ~.
\end{equation}
This equation has $n$ solutions for $\vev{M}_S$. Inserting this into 
the gradient equations
\beq
\pder{\Weff}{g_j}=\frac{1}{j}\vev{M}_S^j
\end{equation}
one can solve for $\Weff$. To see that these gradient equations are 
integrable, we have to show, that there is no curl in the flow. First note
\beq
0=\pder{}{g_k}\left(\sum_jg_j\vev{M}_S^j-S\right)=
\left(\sum_jjg_j\vev{M}_S^{j-1}\right)\pder{\vev{M}_S}{g_k}+\vev{M}_S^k ~.
\end{equation}
Using this we get
\beq
\pder{}{g_k}\frac{1}{j}\vev{M}_S^j=-\frac{\vev{M}_S^{j+k-1}}
{\sum_llg_l\vev{M}_S^{l-1}} ~.
\end{equation}
This shows that the flow is integrable. The integral is again fixed up to a 
function only of $S$, which has to be fixed by asymptotic behavior. Here we 
have $n$ asymptotic regions. One asymptotic region has unbroken $SU(N_c)$ 
gauge group, {\it i.e.} $N_c$ confining vacua. In each of the other asymptotic 
regions the gauge group is Higgsed down to $SU(N_c-1)$, {\it i.e.} there are  
$N_c-1$ vacua, giving rise to a total of $n(N_c-1)+1$ vacua.  

\itsubsub{The Massless Limit Revisited}

We can now use a more general tree level superpotential to calculate the
effective superpotential for the massless case. We use a technique
that will be crucial in dealing with dynamical supersymmetry breaking and
with chiral models where no mass term is possible.

Our aim is to calculate the effective superpotential for
a tree level superpotential
\begin{equation}
W_{tree}=\lambda M^2 ~.
\end{equation}
One possibility in this model is to add a mass term,
apply our technique and then send $m \to 0$ as we have already
done in a previous section.
The other possibility which  is applicable also for chiral models, is
to add a tree level term which gives a classical vacuum
where the gauge group is Higgsed.
We will take
\begin{equation}
W_{tree}=\lambda M^2+\alpha M^4 ~.
\end{equation}
The classical vacua are then $M^2=0$ and
$M^2=-\lambda / 2 \alpha$.
If we solve the Konishi relations as usual we get (for the Higgsed branch)
\begin{eqnarray}
W_{eff}&=&-\frac{\lambda^2}{8 \alpha}- \frac{\lambda^2}{8 \alpha}
\sqrt{1+\frac{4 \alpha}{\lambda^2}S}+1/2 S \log{\alpha}- \nonumber \\
&&-1/2S\log{\lambda}-1/2S\log{(1+ \sqrt{1+\frac{4 \alpha}{\lambda^2}S})} 
+C(S) ~.
\end{eqnarray}
On the Higgsed branch we can easily match it to the 
%perturbative superpotential 
 Veneziano-Yankielowicz potential for $SU(N_c-1)$, whereas on the un-Higgsed
branch a matching 
%with the perturbative calculation 
seems impossible
due to the massless flavors.

Introducing the strong gauge dynamics, we fix the full 
effective superpotential to
\begin{eqnarray}
W_{eff}&=&S \log{\Lambda^{3N_c-1}}-\left(N_c-1 \right)S \log S
+\left(N_c-1 \right)S- \nonumber \\ 
&&-\frac{\lambda^2}{8 \alpha} \mp \frac{\lambda^2}{8 \alpha}
\sqrt{1+\frac{4 \alpha}{\lambda^2}S} 
+\frac{1}{2}S\log{\alpha}
-\frac{1}{2}S\log{\lambda}- \nonumber \\
&&-\frac{1}{2}S\log{\left(\mp \sqrt{1+\frac{4 \alpha}{\lambda^2}S}-1 \right)}
+S \log{2} + \frac{S}{4}  ~,
\end{eqnarray}
where the upper sign corresponds to the Higgsed branch
and the lower sign to the un-Higgsed one.

We want to recover the effective superpotential for the
case $\alpha \to 0$. This limit is not sensible
on the Higgsed branch since 
%the 
its vacua run off to infinity.
The crucial ingredient is that the full superpotential also
knows about the un-Higgsed branch, so we can just change the branch
and take the limit $\alpha \to 0$ there. 
If we do that we indeed recover the result (\ref{Massl1}).
This approach will be used later for the chiral model and models
with dynamical supersymmetry breaking. 

\section{Gauge Group $G_2$ with Three Flavors}\label{sec:g2}

In this subsection we will study \nOne \ SQCD with exceptional
gauge group $G_2$.
We will concentrate on the case with three flavors in the real 
fundamental ${\bf 7}$ representation. This case is instructive
because it requires the introduction of a baryon operator in
addition to mesons, it has an instanton generated superpotential
\cite{Pesando:1995bq,Giddings:1995ns} and exhibits an interesting vacuum
structure \cite{smilga}.

\subsection{The Effective Superpotential}
Using the primitive invariants of
$G_2$ we can construct seven gauge invariant operators. 
Six of them correspond to mesons
\begin{equation}
X_{IJ} = \delta^{ij} Q^i_I Q^j_J ~,
\end{equation}
where $X_{IJ}$ is a symmetric matrix, 
and the seventh operator is the baryon
\begin{equation}
Z = \psi^{ijk} \epsilon_{IJK} Q^i_I Q^j_J Q^k_K ~,
\end{equation}
where 
$\psi^{ijk}$ is the $G_2$ invariant three-tensor which also 
appears in the multiplication table of imaginary octonions.
Note that capital letters, $I,J,K=1,2,3$, denote flavor indices 
whereas small letters, $i,j,k=1,\ldots,7$, denote gauge indices.

For vanishing tree level superpotential the classical theory possesses
a $U(3)$ flavor symmetry.
For concreteness we will study the theory in the presence of
the tree level superpotential
\begin{equation}
\Wtree = m^{IJ} X_{IJ} + \lambda Z ~.
\end{equation}
By a flavor rotation we can always make the mass matrix
$m^{IJ}$ diagonal, but as a further simplification we assume
that all masses are equal 
\begin{equation}
m_{IJ} = m \delta_{IJ}
\end{equation}
which leaves a $SO(3)$ flavor symmetry unbroken. 
From here on we will, therefore, 
consider the $I,J$ indices as $SO(3)$ indices.

Let us now analyze the extrema of the model with this tree-level
superpotential at the classical level. The F-term constraints read
\begin{equation}\label{ftermg2}
Q^i_I = -\frac{3 \lambda}{2 m} \psi^{ijk} \epsilon_{IJK} Q^j_J Q^k_K ~,
\end{equation}
whereas the D-term constraints are
\begin{equation}\label{dtermg2}
\delta_{I J} \overline{Q}^i_{I} T^a_{ij} Q^j_J = 0 ~.
\end{equation}
The $T^a$ are generators of $G_2$ in the fundamental representation.
They furnish a subset of the $SO(7)$ generators and hence are
anti-symmetric in $i,j$.  
There are two solutions to (\ref{ftermg2}) and (\ref{dtermg2}): 
in the first
$Q^i_I = 0$ and the gauge symmetry is unbroken, in the second
$Q^i_I = - \frac{2 m}{3 \lambda} \delta^i_I$, after flavor and 
gauge transformations, which leaves an $SU(2)$ gauge symmetry
unbroken. In the semi-classical regime the matter fields are all
heavy and can be integrated so that the total number of quantum
vacua is the sum of the Witten indices of $G_2$ and $SU(2)$
SYM. This means that we expect six confining vacua with broken
chiral symmetry.

In order to construct the Konishi anomaly relations and its 
generalizations we have to consider two kinds of transformations: 
\begin{equation}
Q^i_I \to Q^i_I + \epsilon \ \lambda_I^J Q^i_J~,
\end{equation}
and
\begin{equation}
Q^i_I \to Q^i_I + \epsilon \ \epsilon_{IJK} \psi^{ijk} Q^j_J Q^k_K ~.
\end{equation}
By contraction of the variations with $\partial W/\partial Q^i_I$ we
obtain the tree-level contributions to the Konishi anomaly which can
be expressed as classical constraints for the gauge invariant meson and 
baryon operators
\begin{equation}
2 m  X^{IJ} + \lambda \delta^{IJ} Z = 0 ~,
\end{equation}
and
\begin{equation}
6 \lambda ( (X^I_I)^2 - X^{IJ} X_{IJ}) + 2 m Z = 0 ~.
\end{equation}
In these variables the two vacuum solutions turn out to be
\begin{equation}
X^{IJ}=Z=0 ~,
\end{equation}
and
\begin{equation}
X^{II} = m^2/\lambda^2 ~,~ X^{IJ} = 0 ~ \mathrm{for} ~ I \neq J 
~,~ Z = -2 m^3/\lambda^3 ~.
\end{equation}

Taking into account the one-loop exact correction to the
Konishi anomaly we find\footnote{
Note that the 2 in front of the gluino operator $S$ is due to the 
fact that the index of the fundamental $7$ representation of $G_2$ is 2,
whereas the fundamental of $SU(N)$ has index 1.
}
\begin{eqnarray}
2 m X^{IJ} + \lambda \delta^{IJ} Z = 2 \ S ~, \nonumber \\
6 \lambda ( (X^I_I)^2 - X^{IJ} X_{IJ}) + 2 m Z = 0 ~. 
\end{eqnarray}
In particular the second line, which corresponds to a generalized
Konishi anomaly, does not receive quantum corrections in this particular
case. The two solution to these quantum relations are:
\begin{equation}
X^{IJ} = x \delta^{IJ} ~,~ Z = -18 \frac{\lambda}{m} x^2
~,~ x_\pm = - \frac{1}{18 \lambda^2} \left( -m^2 \pm
\sqrt{m^4 - 36 \lambda^2 m S} \right) ~,
\end{equation}
where the $x=x_+$ corresponds to the classical vacuum with vanishing
vevs and $x=x_-$ corresponds to the Higgsed vacuum with non-zero vevs.
Using 
\begin{equation}
\frac{\partial \Weff}{\partial m} = 3 \vev{x}_S
~~ \mathrm{and} ~~  \frac{\partial \Weff}{\partial \lambda} 
= \vev{Z}_S = -18 \frac{\lambda}{m} \vev{x}_S^2 ~,
\end{equation}
we can solve for the perturbative part of the effective superpotential
\begin{eqnarray}
\Weff & = &
\frac{m^3}{18 \lambda^2} \left(1 \mp \sqrt{1 - \frac{36 \lambda^2}{m^3} S } 
\right) + \nonumber \\
 & & 2 S \log \left(1 \pm \sqrt{1 - \frac{36 \lambda^2}{m^3}S } \right) + 
%3 S \log \frac{m^3}{\Lambda_0^3} + C(S) .
3 S \log{m^3} + C(S) .
\end{eqnarray}

Matching this in the UV to the $G_2$ theory with three fundamental
chiral multiplets we can fix $C(S)$. We find
\begin{eqnarray}\label{weff:g2}
\Weff & = &
4 S \left( - \log \frac{S}{\hat\Lambda^3} + 1 \right) +
\frac{m^3}{18 \lambda^2} \left(1 - \sqrt{1 - \frac{36 \lambda^2}{m^3}S} 
\right) - \nonumber \\ & &
S + 2 S \log \left(\frac{1}{2} + \frac{1}{2}
\sqrt{1 - \frac{36 \lambda^2}{m^3}S} \right) ~,
\end{eqnarray}
where $\hat\Lambda^{12} = \Lambda^9 m^3$. Looking at the leading $S \log S$ 
terms for the two possible branches of the square roots in (\ref{weff:g2}) 
we find $4+2$ extrema, hence, there are six vacua
as expected \cite{smilga}.

\subsection{The Vacuum Structure}

The quantum vacuum manifold is described by the extrema of
(\ref{weff:g2}). Combining both branches one is led to the
polynomial equation
\begin{equation}\label{qs:g2}
\left(\hat S^3 + \hat S + \frac{z}{2}\right)
\left(\hat S^3 - \hat S + \frac{z}{2}\right)
 = 0~,\quad \hat S = \frac{S}{\hat\Lambda^3}~,
\quad z = \hat\Lambda^3 \frac{18 \lambda^2}{m^3} ~.
\end{equation}
The complex surface defined by equation (\ref{qs:g2}) has six sheets
and there are branch points at $z=\infty$ and 
at the roots of $z^4=256/729$.

\begin{figure}[htb]
\epsfxsize=2.5in
\bigskip
\centerline{\epsffile{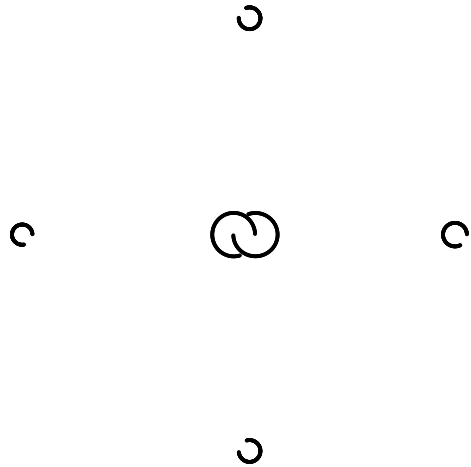}}
\caption{\footnotesize The trivial 
monodromy around $z=0$ as seen in the $\hat S$ plane. The un-Higgsed
vacua, corresponding to the smaller outer circles, are not
interchanged, whereas the two circles in the middle correspond to the
two Higgsed vacua which move around each other but come back to
itself.}
\end{figure}

In order to understand the physics near the critical points we study
the monodromies around them. The situation near $z=0$ is depicted in
Fig.~3. More precisely we are taking a small loop around $z=0$ and
see, contrary to the case of SQCD, that the monodromy is trivial. The
four un-Higgsed vacua of the unbroken $G_2$ theory correspond to the
four outer vacua and stay where they are. The two vacua in the middle
of the picture, which correspond to the vacua where $G_2$ is Higgsed
to $SU(2)$
\footnote{Generically this $SU(2)$ theory confines and has two vacua,
but we call them Higgsed vacua here to distinguish them.}, loop around
each other but eventually return to their starting positions.  In the
limit $z \to 0$ the gluino condensate 
$S$ goes to zero, but this is not related to the 
appearance of a chirally symmetric vacuum. Actually, the two vacua show
run-away behavior as can be seen by inspecting the meson vevs and
$W_{eff}$ which both are driven to infinity in this limit.

More interesting are the critical points located at the roots of
$z^4=256/729$. In Fig.~4 it can be seen that three of the un-Higgsed
$G_2$ vacua (outer circles) and one of the Higgsed vacua (circle in
the center) remain at their original location. However, in the big
circle on the right hand side of the picture we see the exchange of
one Higgsed with one un-Higgsed vacuum.

\begin{figure}[htb]
\epsfxsize=2.5in
\bigskip
\centerline{\epsffile{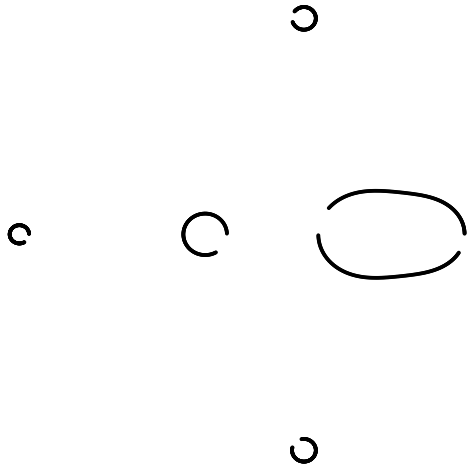}}
\caption{\footnotesize 
Monodromy around $z = 4/\sqrt{27}$ as seen in the $\hat S$ plane. The
un-Higgsed vacua correspond to the smaller outer circles, whereas the
bigger inner circle corresponds to a Higgsed vacuum.  Most interesting
is the rightmost big circle in which we can see the exchange of a
un-Higgsed with a Higgsed vacuum.  }
\end{figure}

Finally, we discuss the monodromy at $z = \infty$. For large $z$ the
six confining, chiral symmetry breaking vacua arrange themselves
symmetrically on a circle. However, the monodromy $z \to z e^{2 \pi
i}$ does not exchange the vacua in a $Z_6$ symmetric fashion but
rather acts like a $Z_3$ rotation on two groups of three vacua, {\it
i.e.} vacua are simultaneously exchanged in the sequence $(1 \to 3 \to
5 \to 1)$ and $(2 \to 4 \to 6 \to 2)$. This behavior can also be
anticipated by the fact that the equation for the quantum parameter
space (\ref{qs:g2}) factorizes. To summarize, the combined actions of
the monodromies permute the six vacua which are organized in two
groups of three but does not lead to exchanges between the two groups.

\section{A Chiral $SU(6)$ Model}\label{sec:ciralmodels}

In this section we apply the developed methods to chiral theories.  We
will start with a model that has well-defined supersymmetric ground
states at the quantum level, and defer the study of the interesting
case of dynamical supersymmetry breaking to a later section.

For concreteness we consider the case of $SU(6)$ with two
antifundamentals $\bar{Q}^I$ and one antisymmetric tensor $X$
\cite{Meurice:1984ai,ads3,ads4,Amati:ft}.  The relevant gauge
invariant operators are
\begin{equation}
T=\varepsilon_{IJ}\bar{Q}_i^I\bar{Q}_j^J X^{ij} ~,
\end{equation}
and
\begin{equation}
U=PfX=X^{i_1j_1}X^{i_2j_2}X^{i_3j_3}\varepsilon_{i_1j_1i_2j_2i_3j_3} ~,
\end{equation}
where the capital $I,J=1,2$ denote flavor indices and the small $i,j$ denote 
color indices. We want to find the effective superpotential for
a theory with the following tree level superpotential 
\begin{equation}
\Wtree=hT+gU ~.
\end{equation}
The classical vacuum is given by $T=U=0$. Since this is a chiral model
we cannot introduce mass terms for the matter fields. We thus cannot
separate the gauge dynamics from the dynamics of the light fields at a
perturbative level and the matching to a one loop calculation will in
general fail. Therefore, we try the technique we have already applied
successfully to the massless SQCD case in section
\ref{sec:gradient}. A deformation of the tree level superpotential
will give a classical vacuum where the gauge group is Higgsed. If all
the light matter degrees of freedom are eaten up by the Higgs
mechanism we can reliably match the effective superpotential below the
Higgsing scale to a one loop calculation in the remaining gauge
group.

We will argue that the following deformation of the tree level
superpotential will give us a classical vacuum with the desired
properties
\begin{equation}
\Wtree = h T + g U + \lambda T U ~.
\end{equation}
The classical vacua have to satisfy the F-flatness conditions
\begin{eqnarray}
h T + \lambda T U & = & 0 ~, \nonumber \\
h T + 3 g U + 4 \lambda T U & = & 0 ~,
\end{eqnarray}
which have two solutions $T=U=0$ and $T=-\frac{g}{\lambda}$, 
$U = -\frac{h}{\lambda}$.
In addition we have to satisfy the D-flatness conditions
\begin{equation}
\bar {Q^\dagger}^j_I \bar Q^I_i - 
\frac{\delta^j_i}{6} \bar {Q^\dagger}^k_{K} \bar Q_k^{K} = 
2 {X^\dagger}_{ik} X^{kj} - \frac{\delta^j_i}{3}{X^\dagger}_{kl} X^{lk} ~,
\end{equation}
where we have taken into account that the $SU(6)$
generators are traceless.
Using gauge and $SU(2)$ flavor rotations we can parametrize the 
solution to these equations as follows:
\begin{equation}
\bar Q = \left( \begin{array}{cccccc} \rho & 0 & 0 & 0 & 0 & 0  \\ 
0 & \rho & 0 & 0 & 0 & 0 \end{array} \right) ~,~
X =  \left( \begin{array}{cccccc} 0 & \lambda_1 & 0 & 0 & 0 & 0  \\
-\lambda_1 & 0 & 0 & 0 & 0 & 0 \\ 
0 & 0 & 0 & \lambda_2 & 0 & 0 \\
0 & 0 & -\lambda_2 & 0 & 0 & 0 \\
0 & 0 & 0 & 0 & 0 & \lambda_2 \\
0 & 0 & 0 & 0 & -\lambda_2 & 0 \\
\end{array} 
\right) ~,
\end{equation}
where D-flatness requires $|\rho|^2 = 2 |\lambda_1|^2 - 2 |\lambda_2|^2$.

From this we can see easily that the gauge group is broken to
$Sp(2)$. First the $\bar Q$ break $SU(6) \to SU(4)$ and the subgroup
of $SU(4)$ that leaves invariant the lower four-by-four block of $X$
is $Sp(2)$.  At the same time 25 out of the 27 matter fields are eaten
up by the Higgs mechanism. The two remaining matter fields are massive
singlets under the $Sp(2)$.  Hence, at low energies the theory becomes
pure glue.

The anomalous Konishi variations give
\begin{eqnarray}
hT+\lambda TU&=& S ~, \nonumber \\
hT+3gU+4\lambda TU&=&4 S ~,
\end{eqnarray}
with $S=-\frac{1}{32\pi^2}\tr_f W_\alpha W^\alpha$ such that we have
\begin{eqnarray}
\label{ChK1}
T&=&\frac{g}{2\lambda}\left(-1\pm\sqrt{1+\frac{4\lambda}{gh}S}\right)~,
\nonumber \\
U&=&\frac{h}{2\lambda}\left(-1\pm\sqrt{1+\frac{4\lambda}{gh}S}\right)~.
\end{eqnarray}

The perturbative part of the effective superpotential can then again
be determined by using gradient equations, it turns out to be
\beq
\Weff=\frac{gh}{2\lambda}\left(-1\pm\sqrt{1+\frac{4\lambda}{gh}S}\right)+
S\log gh+2S\log\left(\half\pm\half\sqrt{1+\frac{4\lambda}{gh}S}\right)+C(S)
~.
\end{equation}
As can be seen from (\ref{ChK1}) the upper $(+)$ sign corresponds
to the un-Higgsed branch, whereas the lower $(-)$ sign corresponds
to the Higgsed branch.

 To determine $C(S)$ we match the effective superpotential
to the Veneziano-Yankielowicz potential of the (classically) 
unbroken gauge group on the Higgsed branch.
%To determine $C(S)$ we match the effective superpotential
%to a perturbative calculation on the Higgsed branch.
We know that on the Higgsed branch below the Higgsing scale
the theory is described by pure $Sp(2)$ glue.
If we take for the purpose of matching the bare parameters
$g=h$ then the unique Higgsing scale is $\Delta=T^{1/3}=U^{1/3}
=(-g/\lambda)^{1/3}$. 
%A perturbative calculation around the 
%Higgsed vacuum gives
%\begin{equation}
%\label{perti}
%W_{eff}^{pert.}=9S \log{\frac{\Lambda_2}{\mu}} +O(S^2) \ ,
%\end{equation}
 The strong coupling dynamics
of $Sp(2)$ are described by
\begin{equation}
\label{perti}
W_{VY}=3S \left(-\log{\frac{S}{\Lambda_2^3}}+1 \right) \ ,
\end{equation}
where the scale $\Lambda_2$ of the $Sp(2)$ is related to the scale
$\Lambda_6$ 
of the $SU(6)$ by the usual matching of scales at the Higgsing scale
$\Delta$ as
\begin{equation}
\left(\frac{\Lambda_6}{\Delta}\right)^{15}=
\left(\frac{\Lambda_2}{\Delta} \right)^9 \ .
\end{equation}
Demanding that the effective superpotential reproduce the 
%perturbative result 
potential (\ref{perti}) below $O(S^2)$ we fix the $C(S)$. 
%Introducing the
%strong coupling gauge dynamics as usual by replacing the IR cutoff
%$\mu$ as $\mu^3 \to S/e$ 
We find the full effective superpotential
\begin{eqnarray}
\label{fully1}
W_{eff} & =& -5S\log{\frac{S}{\Lambda_6^3}} + 4S -2S \log{2}+ \nonumber \\
&& + S\log{gh}+\frac{gh}{2\lambda}\left(-1\pm
\sqrt{1+\frac{4\lambda}{gh}S}\right)+ \nonumber \\
&& + 2S\log\left(\half\pm\half\sqrt{1+\frac{4\lambda}{gh}S}\right) \ .
\end{eqnarray}
If we want to recover the superpotential for the theory with
$\lambda=0$ we have to go to the un-Higgsed branch which corresponds
to the upper $(+)$ sign.  We emphasize again that we have used the
semiclassical region of the  Higgsed branch to determine the
integration ``constant'' $C(S)$, but once we have obtained the full
effective superpotential we can use its analytic structure and move
freely among the branches.

In the $\lambda \to 0$ limit on the un-Higgsed branch we thus
obtain
\begin{equation}
W_{eff}=5S\left(-\log{\frac{S}{\Lambda_6^3}}+1 \right)+S\log{gh} ~.
\end{equation}
This effective superpotential precisely reproduces the 
5 vacua which were found  e.g. in 
\cite{Amati:ft} by instanton calculations
\begin{equation}
S=\Lambda_6^3\left(gh \right)^{\frac{1}{5}}e^{2\pi i k/5} ~.
\end{equation}

\section{Dynamical SUSY Breaking}\label{sec:dsb}

We now want to see how the chiral ring and the Konishi anomaly can be used 
to understand theories with dynamical supersymmetry breaking. Naively,
we cannot apply the method we used in the previous sections, since 
%$|\rangle_S$
there is no supersymmetric vacuum.
%and factorization of expectation values of chiral operators does not apply. 
However, one can deform the original
tree level superpotential to get supersymmetric vacua in which we can make
a reliable calculation of the effective superpotential.
One can then analyze the behavior of the 
vacua when switching the deformation off again.

To be specific we discuss a variant of the 
Izawa-Yanagida-Intriligator-Thomas (IYIT) model \cite{Izawa:1996pk,intrili}
which features dynamical supersymmetry breaking (DSB). 
The model is ${\cal N}=1$ supersymmetric $Sp(N_c)$ gauge theory with 
$2N_f=2(N_c+1)$ fundamental chiral multiplets $Q_a^i$, ($a=1,...,2N_c,\; 
i=1,...,2N_f$),
and a gauge singlet chiral multiplet $S_{ij}$, which is antisymmetric in the 
indices $i,j$. The gauge invariant matter fields of this theory are the meson 
$M^{ij}=Q^iQ^j$ and $S_{ij}$.
Note that we will denote both the gauge singlet and the
glueball superfield by $S$. However, the former always carries flavor 
indices, so no confusion should arise.
We consider the above theory with a tree level potential given by
\begin{equation}
\Wtree=\lambda S_{ij}M^{ij} -m J^{ij}S_{ij} \ ,
\end{equation}
where $J=\Bid_{N_f}\otimes i\sigma^2$ is the symplectic form.
This model has been studied in  \cite{deBoer:1998by}.
The effective superpotential is 
\begin{equation}
\label{eq:Superpot1}
\Weff= X\left( Pf \, M - \Lambda^{2N_c+2} \right)+ \lambda S_{ij}M^{ij} 
-m J^{ij}S_{ij} \ .
\end{equation}
Integrating out the Lagrange multiplier field $X$ and the gauge singlet
$S_{ij}$ we get\footnote{The Pfaffian of $M$ is defined here as 
$Pf\,M=\frac{1}{2^N_f N_f!}\epsilon_{i_1j_1,...,j_{N_f}}
M^{i_1j_1}...M^{i_{N_f}j_{N_f}}$, 
such that $Pf\,J=1$.}
\begin{equation}
Pf \, M = \Lambda^{2N_c+2}, \; \; M=\frac{m}{\lambda}J \ .
\end{equation}
From these equations it follows that the above superpotential has a 
minimum only when
\begin{equation}
\label{eq:Condi1}
\left(\frac{m}{\lambda}\right)^{N_c+1}= \Lambda^{2N_c+2} \ .
\end{equation}
This is the condition on the bare parameters for unbroken supersymmetry.

\subsection{Massive Deformation}
% and Dynamical Supersymmetry Breaking}

We now turn to the problem to derive (\ref{eq:Condi1}) without knowing
the superpotential (\ref{eq:Superpot1}).  
%A necessary condition for
%the dynamical supersymmetry breaking was the absence of a classical
%supersymmetric vacuum with only massive matter. 
In order to guarantee
the existence of a supersymmetric vacuum, we can deform the tree level
superpotential in such a way that there is a classical vacuum in which
all matter becomes massive.  This allows to implement the strong IR
dynamics and the full quantum theory has supersymmetric vacua. When
turning those deformations off, we will see, that all the vacua run
away, unless (\ref{eq:Condi1}) is satisfied.

We deform the above theory by giving the gauge singlets $S_{ij}$ 
a mass $\alpha S_{ij}S^{ij}$. The tree level superpotential is 
\begin{equation}
W_{tree}=\lambda S_{ij}M^{ij} -m J^{ij}S_{ij}+\alpha S_{ij}S^{ij} \ ,
\end{equation}
where $S^{kl}=S_{ij}J^{ik}J^{jl}$. This potential has many vacua. 
Two of them will be important in the following. 
The vacuum with all squarks massless at $M^{ij}=\frac{m}{\lambda}J^{ij}$ and $S_{ij}=0$. 
It exists also for $\alpha=0$. 
The other one is the massive vacuum, with $Q=0$ and $S_{ij}=-\frac{m}{2\alpha}(J^{-1})_{ij}$.
Around the second classical vacuum we can integrate out the massive fields, such 
that in the IR we are left with pure gauge theory. 

To obtain the Konishi relations we consider the variations
$\delta_1 Q^i=\epsilon^i_j Q^j$, and $\delta_2 S_{ij}= 
\epsilon_{ij}\;^{lm} S_{lm}$. These give rise to the following 
respective relations
\begin{eqnarray}
\label{eq:K1}
&& 2\lambda \vev{S_{ij}M^{kj}}_S = \delta^k_i S  ~, \nonumber \\
&& \lambda \vev{S_{ij}M^{kl}}_S-m J^{kl}\vev{S_{ij}}_S +2 
\alpha \vev{S^{kl}}_S\vev{S_{ij}}_S  =  0  ~. 
\end{eqnarray}

After straightforward algebra one can solve for the expectation values of 
the chiral operators,
\begin{eqnarray}
\label{solsusbr}
S_{ij}&=&(A\otimes i\sigma^2)_{ij} ~, \quad 
A_{ab}= \delta_{ab} \left( \frac{m}{4\alpha}-\eta_a \sqrt{
\frac{m^2}{16\alpha^2}-\frac{S}{4\alpha}} \right) ~, \quad 
a,b=1,...,N_f ~, \nonumber \\
M^{ij}&=&(mJ^{ij}-2\alpha S^{ij})/\lambda ~,
\end{eqnarray}
where we used flavor rotations to bring $A$ to diagonal form and
$\eta_a$ denotes a choice of a vector with components $\pm 1$. The
number of $\pm 1$ entries in $\eta_a$ will be denoted by $N_f^\pm$.
 Note that $N_f^+=0$ corresponds to the massive vacuum, as can be seen 
in the classical limit of $S$ small. $N_f^+=N_f$ corresponds to 
the classical vacuum, which exists also for $\alpha=0$.

The expectation values of the chiral composites then read,
\begin{eqnarray}
\label{koneqnIYIT}
\vev{S_{ij}M^{ij}}_S & = & N_f \frac{S}{\lambda} ~, \\ \nonumber
\vev{J^{ij}S_{ij}}_S& = & N_f^+\frac{m}{\alpha}\left(\frac{1}{2}-
\frac{1}{2}\sqrt{1-\frac{4\alpha S}{m^2}}\right)+N_f^-\frac{m}{\alpha}
\left(\frac{1}{2}+\frac{1}{2}\sqrt{1-\frac{4\alpha S}{m^2}}\right)~, \\ 
\nonumber
\vev{S_{ij}S^{ij}}_S& = & N_f^+\frac{m^2}{2\alpha^2}\left(\frac{1}{2}-
\frac{1}{2}\sqrt{1-\frac{4\alpha S}{m^2}}\right)^2+N_f^-\frac{m^2}{2\alpha^2}
\left(\frac{1}{2}+\frac{1}{2}\sqrt{1-\frac{4\alpha S}{m^2}}\right)^2
\nonumber ~.
\end{eqnarray}
Integrations with respect to the various parameters then 
gives the perturbative part 
of the effective superpotential $\Weff^{pert}$. By matching 
$\Weff^{pert}$ to the VY potential, which describes the pure 
gauge dynamics around the classical vacuum
$Q=0$ and $S_{ij}=-\frac{m}{2\alpha}(J^{-1})_{ij}$, gives the following
effective superpotential,
\begin{eqnarray}\label{DSBWeff}
\Weff&=&(N_f)\left[S\log{\frac{\Lambda^3}{S}}+1\right]+N_f S 
\log{\left(\frac{\lambda S}{\Lambda m}\right)}
-N_f\left(\frac{m^2}{4\alpha}+\frac{S}{2}\right)\nonumber\\&&
+(N_f^+-N_f^-)\frac{m^2}{4\alpha}\sqrt{1-\frac{4 \alpha}{m^2}S}-\nonumber\\
&&-S\log{\!\!\left[\left(\frac{1}{2}+\frac{1}{2}\sqrt{1-\frac{4 \alpha}{m^2}S}
\right)^{N_f^+}\!\!\left(\frac{1}{2}-\frac{1}{2}\sqrt{1\!-\!\frac{4 \alpha}{m^2}S}
\right)^{\!N_f^-}\right]} .
~~\end{eqnarray}
The derivative $\partial_S\Weff(S)=0$ then leads to
\begin{eqnarray}
  \label{eq:vaceqweff}
  \log{\left[\left(\frac{m}{\Lambda^2\lambda}\right)^{N_f}\!r
\!\!\left(\frac{1}{2}+
\frac{1}{2}\sqrt{1\!-\!\frac{4 \alpha}{m^2}S}
\right)^{N_f^+}\!\!\left(\frac{1}{2}-\frac{1}{2}\sqrt{1\!-\!\frac{4 \alpha}{m^2}S}
\right)^{N_f^-}\right]}\!\!=\!0 ~.
\end{eqnarray}
For $N_f^-=N_f$ this can be solved explicitly to give
\begin{eqnarray}\label{eq:expsolDSB}
S&=&e^{2\pi ik/(N_c+1)}\Lambda^2\frac{m\lambda}{\alpha}
\left(1-e^{2\pi ik/(N_c+1)}\frac{\Lambda^2\lambda}{m}\right) ~,
\nonumber \\
S_{ij}&=&-\frac{m}{2\alpha}\,\left(1-e^{2\pi ik/(N_c+1)}
\frac{\Lambda^2\lambda}{m}\right)(J^{-1})_{ji} ~, \nonumber \\
M^{ij}&=&e^{2\pi ik/(N_c+1)}\,\Lambda^2\, J^{ji} ~.
\end{eqnarray}
In total we find $N_c+1$ vacua for each point of the parameter space. 
The massive vacuum can be found in the limit of small $\Lambda^2\lambda/m$. 
The classical vacuum which exists also for $\alpha=0$ corresponds
to the points $m/\lambda\Lambda^2=e^{2\pi ik/(N_c+1)}$.

%We will discuss this result further in certain limits. 
%Let us first consider the case $\Lambda^2\lambda/m\ll1$. 
%The expectation value of $S_{ij}$ is close 
%to the classical value $S_{ij}=-\frac{m}{2\alpha}(J^{-1})_{ij}$. 
%The gauge singlets $S_{ij}$ and the quarks $Q$ are massive and 
%integrated out. Thus we encounter pure $Sp(N_c)$ 
%gauge dynamics with the scale 
%$\hat\Lambda^{3(N_c+1)}=\Lambda^{2(N_c+1)}m_Q^{N_c+1}$. 
%Here the masses of the charged fields $Q$ enter the scale $\hat\Lambda$. 
%Remember that 
%$m_Q=\frac{\lambda m}{\alpha}$. It is instructive to consider also $\Weff$ 
%given in (\ref{DSBWeff}) in this limit. One finds the expected VY potential. 

%For a large mass scale $\alpha$ the expectation values of $S$ 
%and $S_{ij}$ become large, 
%such that the $\alpha\rightarrow 0$ limit does not give well 
%defined vacua for $\Lambda^2\lambda/m=0$. 

For generic points in the parameter space with 
$N_f^-$ arbitrary the zero mass limit gives run-away vacua. 
Only for $N_f^-=N_c+1$ and $m/\lambda\Lambda^2=e^{2\pi ik/(N_c+1)}$ 
we find finite expectation values in the $\alpha\rightarrow 0$ limit.
The vacuum with $S=S_{ij}=0$ and 
$M^{ij}=e^{2\pi ik/(N_c+1)}\Lambda^2\,J^{ij}$ stays finite, whereas the $N_c$
other vacua still run away.
We recover the quantum constraint on the parameters (\ref{eq:Condi1}).

 As is by now familiar
we expect to recover
(\ref{eq:Condi1}) directly if we calculate the effective superpotential
for the theory with $\alpha=0$ by switching branches and taking the limit
$\alpha \to 0$. The other branch corresponds to $N_f^+=N_f$
and sending $\alpha$ to zero gives
\begin{equation}
\label{HPot1}
W_{eff}=N_f S \log{\frac{\Lambda^2 \lambda}{m}} \ .
\end{equation}
 It is interesting that $S$ appears just as a Lagrange multiplier
for the constraint (\ref{eq:Condi1}) in this superpotential.
A similar appearance of $S$ in the effective superpotential
was observed in \cite{Bena:2002ua}.

This method can be generalized to other models with dynamical 
supersymmetry breaking. A mass term will typically produce a supersymmetric 
quantum vacuum. In the limit of turning off the mass, one can 
see how the quantum vacua run away, except for some vacua, which might 
stay finite for certain choices of the other parameters.
We can see here the mechanism by which the Witten index jumps, when the 
highest couplings in the tree level superpotential are switched off.

\subsection{An Alternative Derivation}

In this section we will derive (\ref{eq:Condi1}) using the 
Konishi anomaly without using a mass deformation.
Our strategy will be to assume unbroken supersymmetry, so that
we can use the Konishi anomaly relations and derive an effective 
superpotential as a function of $S$. 
Minimizing this effective superpotential with respect to 
$S$ should lead to (\ref{eq:Condi1}).

To obtain the Konishi relations we consider the variations
$\delta_1 Q^i=\epsilon^i_j Q^j$, and $\delta_2 S_{ij}= 
\epsilon_{ij}\;^{lm} S_{lm}$.
These give rise to the following respective relations
\begin{eqnarray}
\label{eq:K0}
2 \lambda \vev{S_{ij}M^{kj}}_S & = & \delta_i^k S ~, \nonumber \\
\lambda \vev{S_{ij}M^{kl}}_S-m J^{kl}\vev{S_{ij}}_S & = & 0 ~.
\end{eqnarray}
The equations (\ref{eq:K0}) contain a lot of valuable information. As usual,
they enable us to derive the dependence of $\Weff$ on the bare parameters.
Rewriting (\ref{eq:K0}) as
\begin{eqnarray}
\frac{\partial  \Weff}{\partial \lambda} & = & N_f \frac{S}{\lambda} ~, 
\nonumber \\
\lambda \frac{\partial  \Weff}{\partial \lambda} + m \,
\frac{\partial  \Weff}{\partial m} & = & 0 ~,
\end{eqnarray}
we can solve for $\Weff$ to get
\begin{equation}
\label{eq:Dep1}
\Weff(S,\lambda,m) = N_f S \log {\frac{\lambda}{m}} + C(S) ~. 
\end{equation}
However, due to the factorization of the chiral vevs we can also 
rewrite the conditions in (\ref{eq:K0}) as 
\begin{eqnarray}
\lambda \vev{S_{ij}}_S\vev{M^{kj}}_S & = & \delta_i^k S ~, \nonumber \\
\lambda \vev{S_{ij}}_S\vev{M^{kl}}_S-m J^{kl}\vev{S_{ij}}_S & = & 0 ~,
\end{eqnarray}
and solve for $\vev{S_{ij}}_S$ and $\vev{M^{ij}}_S$. We get
\begin{equation}
\label{eq:Condi2}
\vev{M^{ij}}_S=\frac{m}{\lambda}J^{ij}, \;\; 
\vev{S_{ij}}_S=\frac{1}{2}\frac{S}{m}\left(J^{-1}\right)_{ji} ~.
\end{equation}
We have now gathered enough information to turn to the derivation of 
the full $\Weff(S,\lambda,m)$. First, we think of $\lambda S_{ij}$
as a mass for the fundamental chiral multiplets and integrate them out.
This will give us an effective superpotential as a function of 
$(S,S_{ij},\lambda,m)$. 
Note that in this model the canonical mass term for the fundamentals
is $\frac{1}{2} m_{ij}Q^i Q^j$ such that the canonical mass
is expressed as $m_{ij}=2 \lambda S_{ij}$.
A perturbative evaluation then yields
\begin{equation}
\Weff^{pert}=3\left(N_c+1\right)S\log{\frac{\Lambda}{\mu}}
+S\log{Pf\left(\frac{2\lambda}{\Lambda} S_{ij} \right)} + 
W^{(1)}(S,S_{ij},\lambda,m) ~.
\end{equation}
Note that we have already taken into account the
contribution of the bare coupling $\tau$ enabling us to replace the 
UV cutoff by the dynamically generated scale.
The part $W^{(1)}$ will be determined by the requirements that 
(a) the extremal value of $S_{ij}$ satisfy the second 
equation of (\ref{eq:Condi2}) and that (b) the superpotential 
obtained after integrating out $S_{ij}$ have the appropriate dependence 
on the bare parameters (\ref{eq:Dep1}). 
Both requirements uniquely fix the effective action to 
\begin{equation}
\Weff^{pert}=3\left(N_c+1\right)S\log{\frac{\Lambda}{\mu}}
+S\log{Pf\left(\frac{2\lambda}{\Lambda} S_{ij} \right)} -m J^{ij}S_{ij} ~.
\end{equation}
Indeed, after integrating out $S_{ij}$ and replacing the first part 
by the appropriate Veneziano-Yankielowicz term, we find
\begin{equation}
\Weff= \left(N_c+1\right)S \left[\log{\frac{\Lambda^3}{S}}+1 \right]
+ S\log{Pf\left[\frac{\lambda S}{m \Lambda} \left(J^{-1} \right)_{ji}\right]}
-N_f S ~,
\end{equation}
which has the appropriate dependence on the bare parameters.
After taking into account that $N_f=N_c+1$ the effective superpotential 
simplifies to
\begin{equation}
\Weff=\left(N_c+1\right)S\log{\frac{\lambda \Lambda^2}{m}} ~.
\end{equation}
This is the expected expression (\ref{HPot1}).
Minimizing with respect to $S$ gives the relation
\begin{equation}
\left(\frac{m}{\lambda}\right)^{N_c+1}= \Lambda^{2 
\left(N_c+1\right)} ~. 
\end{equation}

\begin{acknowledgments}
The work of A.B. and C.R.\ is supported in part by funds  
provided by the DOE under grant number DE-FG03-84ER-40168.
The work of H.I, H.N. and Y.O.\ is supported in part by the US-Israel
BSF and the TMR network.
We have greatly benefited from discussions with Richard Corrado, 
Jaume Gomis, Nick Halmagyi, Vadim Kaplunovsky,
Kristian Kennaway,   
Krzysztof Pilch, Tadakatsu Sakai, Cobi Sonnenschein and Nicholas Warner.
Y.O. would like to thank the theory group in Caltech for hospitality
during the initial stages of the work. 
\end{acknowledgments}

\newpage 

\begin{appendix}

\section{Proof of One Loop Exactness of the Konishi Anomaly}

In this appendix we want to show how the proof for the one loop exactness of the generalized Konishi anomaly works. We follow the idea of \cite{Cachazo:2002ry}. We concentrate on the normal Konishi anomaly for SQCD with one flavor for concreteness. It is easy to generalize this proof to other cases.

There are two flavor symmetries $U(1)_Q$ and $U(1)_{\tilde Q}$ together with an R-symmetry $U(1)_R$. Those symmetries are broken by the tree level superpotential and by anomalies. By promoting the coupling constants to chiral superfields, which transform under those symmetries, we can restore those symmetries. The charges are summarized in the following table
\beq
\begin{array}{c|ccc}
               & U(1)_Q & U(1)_{\tilde Q} & U(1)_R        \\
\hline
Q              & 1      & 0               & \frac{2}{3}   \\
\tilde Q       & 0      & 1               & \frac{2}{3}   \\
W_\alpha       & 0      & 0               & 1             \\
m              & -1     & -1              & \frac{2}{3}   \\
\lambda        & -2     & -2              & -\frac{2}{3}  \\
\Lambda^{3N-1} & 1      & 1               & 2N-\frac{2}{3} 
\end{array}
\end{equation}

We want to calculate the lowest component of the Konishi anomalies for
\beq\label{konsym}
Q\mapsto Q+\epsilon Q\quad {\rm and}\quad \tilde Q\mapsto \tilde Q+\tilde\epsilon \tilde Q.
\end{equation}
Let us concentrate on the first Konishi anomaly. We want to calculate the 
divergence of the supercurrent associated with the first transformation
in (\ref{konsym})
\beq
\bar D^2J=\bar D^2Q^\dagger e^V Q=\pder{\Wtree}{q}q+\CO(\theta,\bar\theta)
\end{equation}
in a slowly varying background gaugino field. The lowest component of this expression is a chiral operator. This chiral operator depends (modulo $\bar Q$ exact operators) only on other chiral operators and it depends only holomorphically on the coupling constants. Furthermore we assume, a smooth weak coupling behavior, {\it i.e.} the coupling constants can only appear with positive integer powers. The scale $\Lambda$ can only appear in positive integer powers of $\Lambda^{3N-1}$, because the leading non-perturbative effects at weak coupling are due
regular instantons and, in particular, 
we do not expect fractional instantons to contribute. 
We also assume, that all fields can only appear with positive powers, {\it i.e.} 
that there is no singularity at the origin in field space.

The divergence $\bar D^2J$ has the charges $(0,0,2)$. Charge conservation then gives constraints on the powers in which all the fields and coupling constants can appear
\beq\label{eqn:chargecons}
n_Q\vect{1}{0}{\frac{2}{3}}
+n_{\tilde Q}\vect{0}{1}{\frac{2}{3}}
+n_{W_\alpha}\vect{0}{0}{1}+n_m\vect{-1}{-1}{\frac{2}{3}}
+n_\lambda\vect{-2}{-2}{-\frac{2}{3}}
+n_{\Lambda^{3N-1}}\vect{1}{1}{\frac{6N-2}{3}}=
\vect{0}{0}{2}
\end{equation}
The first two equations imply $n_{\tilde Q}=n_Q$. Subtracting four times the first equation from the third equation, we get
\beq
n_{W_\alpha}+2n_m+2n_\lambda+(2N-2)n_{\Lambda^{3N-1}}=2.
\end{equation}
For $N>2$ this implies, that $n_{\Lambda^{3N-1}}=0$, {\it i.e.} there are no nonperturbative contributions. This, together with the first equation of the charge conservation, leaves us with three kinds of solutions
\beq\label{threecases}
\begin{array}{c|cccccc}
   & n_Q & n_{\tilde Q} & n_{W_\alpha} & n_m & n_\lambda & n_{\Lambda^{3N-1}}\\
\hline
1. & 0   & 0            & 2            & 0   & 0         & 0                \\
2. & 1   & 1            & 0            & 1   & 0         & 0                \\
3. & 2   & 2            & 0            & 0   & 1         & 0                
\end{array}
\end{equation}

We now need to determine, which kinds of diagrams can contribute in each of 
those three cases. To this end we need to look at the Feynman rules. Since 
we want to calculate correlators with only chiral fields as external legs, 
which depend holomorphically on the coupling constants, there are only three 
kinds of vertices, that can contribute.
\begin{itemize}
\item The vertex of the current
\beq
\pder{\Wtree}{\phi}\phi ~,
\end{equation}
\item The vertex due to the superpotential
\beq
\frac{\partial^2\Wtree}{\partial\phi}\psi^2 ~
\end{equation}
\item The coupling to the gaugino
\beq
\phi^\dagger W_\alpha\psi^\alpha ~.
\end{equation}
\end{itemize}

The first two kinds of vertices come with a coupling constant, whereas the 
third kind corresponds to the insertion of background gaugino field. 
Therefore, the number of vertices in a diagram is given by
\beq
V=\sum_j n_{g_j}+n_{W_\alpha},
\end{equation}
where the coupling constants are denoted by $g_j$.
The number of propagators can be determined by counting the number of internal legs on those vertices
\beq
P=\half\left(\sum_j n_{g_j}l_j+2n_{W_\alpha}-n_\phi\right),
\end{equation}
where $l_j$ is the number of legs (bosonic and fermionic) on the vertex $j$.
We can combine those two results to get the number of loops $L$ in a diagram\footnote{This is the number of momentum loops, not the number of index loops in a ribbon graph.}
\beq
L=1+P-V=1+\half\left(\sum_j n_{g_j}(l_j-2)-n_\phi\right).
\end{equation}

Inserting the previous results (\ref{threecases}) into this formula we see, 
that there are only tree level and one loop diagrams contributing to the 
lowest component of the Konishi anomaly, {\it i.e.} the anomaly is one loop exact 
and we can trust our expressions. This argument can easily be generalized to 
theories with different gauge groups and matter content, and also to the 
generalized Konishi anomaly. It is easy to see that a sufficient condition
for the one loop exactness of the generalized Konishi anomaly is
\beq\label{eqn:indexcond}
2C(adj)-\sum_I 2C(r_I)>2,
\end{equation}
where the sum is over all matter fields and $2C(r)$ is the index of the
representation $r$. This condition is satisfied in most of the cases we 
study. However, if (\ref{eqn:indexcond}) is not satisfied, 
one needs to study the full set of charge conservation equations, in 
analogy to (\ref{eqn:chargecons}).
Sometimes the one loop exactness can fail, e.g. for too 
small gauge groups or for a 
sufficiently large number of external legs, $\Lambda$ dependent terms can 
appear, which
correspond to non-perturbative corrections to Konishi anomalies. We have not
found an argument for the absence of such terms in general, but for the 
purpose of calculating superpotential such corrections do not appear in the
examples studied in this paper.

\end{appendix}

\vfill
\pagebreak

\providecommand{\href}[2]{#2}\begingroup\raggedright\endgroup

\end{document}